\documentclass[preprint,amsmath,amssymb,pra,superscriptaddress]{revtex4-2}

\usepackage[utf8]{inputenc}
\newcommand{\twoD}{2D$^+$-MOT~}
\newcommand{\Rb}{$^{87}$Rb~}

\usepackage{graphicx}
\usepackage{multirow}
\usepackage{amsmath,amssymb,amsfonts}
\usepackage{amsthm}
\usepackage{mathrsfs}
\usepackage{xcolor}
\usepackage{textcomp}
\usepackage{booktabs}
\usepackage{algorithm}
\usepackage{algorithmicx}
\usepackage{algpseudocode}
\usepackage{listings}

\begin{document}
	
	\title{INTENTAS - An entanglement-enhanced atomic sensor for microgravity}
	
	\author{O. Anton}
	\affiliation{Humboldt Universit\"at zu Berlin, 12489 Berlin, Germany}
	
	\author{I. Br\"ockel}
	\affiliation{Deutsches Zentrum f\"ur Luft- und Raumfahrt e.V. (DLR), Institut f\"ur Satellitengeod\"asie und Inertialsensorik (DLR-SI), 30167 Hannover, Germany}
	
	\author{D. Derr}
	\affiliation{Technische Universit\"at Darmstadt, Fachbereich Physik, Institut f\"ur Angewandte Physik, 64289 Darmstadt, Germany}
	
	\author{A. Fieguth}
	\email{alexander.fieguth@dlr.de}
	\affiliation{Deutsches Zentrum f\"ur Luft- und Raumfahrt e.V. (DLR), Institut f\"ur Satellitengeod\"asie und Inertialsensorik (DLR-SI), 30167 Hannover, Germany}
	
	\author{M. Franzke}
	\affiliation{Leibniz Universit\"at Hannover, Institut für Quantenoptik, 30167 Hannover, Germany}
	
	\author{M. G\"artner}
	\affiliation{Ferdinand-Braun-Institut (FBH), 12489 Berlin, Germany}
	
	\author{E. Giese}
	\affiliation{Technische Universit\"at Darmstadt, Fachbereich Physik, Institut f\"ur Angewandte Physik, 64289 Darmstadt, Germany}
	
	\author{J. S. Haase}
	\email{haase@iqo.uni-hannover.de}
	\affiliation{Leibniz Universit\"at Hannover, Institut für Quantenoptik, 30167 Hannover, Germany}
	
	\author{J. Hamann}
	\affiliation{Leibniz Universit\"at Hannover, Institut für Quantenoptik, 30167 Hannover, Germany}
	
	\author{A. Heidt}
	\affiliation{Leibniz Universit\"at Hannover, Institut für Quantenoptik, 30167 Hannover, Germany}
	\affiliation{Leibniz Universit\"at Hannover, Institut für Transport- und Automatisierungstechnik, 30823 Garbsen, Germany}
	
	\author{S. Kanthak}
	\affiliation{Humboldt Universit\"at zu Berlin, 12489 Berlin, Germany}
	
	\author{C. Klempt}
	\affiliation{Deutsches Zentrum f\"ur Luft- und Raumfahrt e.V. (DLR), Institut f\"ur Satellitengeod\"asie und Inertialsensorik (DLR-SI), 30167 Hannover, Germany}
	\affiliation{Leibniz Universit\"at Hannover, Institut für Quantenoptik, 30167 Hannover, Germany}
	
	\author{J. Kruse}
	\affiliation{Deutsches Zentrum f\"ur Luft- und Raumfahrt e.V. (DLR), Institut f\"ur Satellitengeod\"asie und Inertialsensorik (DLR-SI), 30167 Hannover, Germany}
	
	\author{M. Krutzik}
	\affiliation{Humboldt Universit\"at zu Berlin, 12489 Berlin, Germany}
	\affiliation{Ferdinand-Braun-Institut (FBH), 12489 Berlin, Germany}
	
	\author{S. Kubitza}
	\affiliation{Ferdinand-Braun-Institut (FBH), 12489 Berlin, Germany}
	
	\author{C. Lotz}
	\affiliation{Leibniz Universit\"at Hannover, Institut für Transport- und Automatisierungstechnik, 30823 Garbsen, Germany}
	
	\author{K. M\"uller}
	\affiliation{Deutsches Zentrum f\"ur Luft- und Raumfahrt e.V. (DLR), Institut f\"ur Satellitengeod\"asie und Inertialsensorik (DLR-SI), 30167 Hannover, Germany}
	
	\author{J. Pahl}
	\affiliation{Humboldt Universit\"at zu Berlin, 12489 Berlin, Germany}
	
	\author{E. M. Rasel}
	\affiliation{Leibniz Universit\"at Hannover, Institut für Quantenoptik, 30167 Hannover, Germany}
	
	\author{M. Schiemangk}
	\affiliation{Ferdinand-Braun-Institut (FBH), 12489 Berlin, Germany}
	
	\author{W. P. Schleich}
	\affiliation{Institut für Quantenphysik and Center for Integrated Quantum Science and Technology (IQST), Universität Ulm, 89081 Ulm, Germany}
	
	\author{S. Schwertfeger}
	\affiliation{Ferdinand-Braun-Institut (FBH), 12489 Berlin, Germany}
	
	\author{A. Wicht}
	\affiliation{Ferdinand-Braun-Institut (FBH), 12489 Berlin, Germany}
	
	\author{L. W\"orner}
	\affiliation{Deutsches Zentrum f\"ur Luft- und Raumfahrt e.V. (DLR), Institut f\"ur Satellitengeod\"asie und Inertialsensorik (DLR-SI), 30167 Hannover, Germany}
	
	\begin{abstract}
		The INTENTAS project aims to develop an atomic sensor utilizing entangled Bose-Einstein condensates (BECs) in a microgravity environment. This key achievement is necessary to advance the capability for measurements that benefit from both entanglement-enhanced sensitivities and extended interrogation times. The project addresses significant challenges related to size, weight, and power management (SWaP) specific to the experimental platform at the Einstein-Elevator in Hannover. The design ensures a low-noise environment essential for the creation and detection of entanglement. Additionally, the apparatus features an innovative approach to the all-optical creation of BECs, providing a flexible system for various configurations and meeting the requirements for rapid turnaround times. Successful demonstration of this technology in the Einstein-Elevator will pave the way for a future deployment in space, where its potential applications will unlock high-precision quantum sensing.
	\end{abstract}
	
	\maketitle

\section{Introduction}

Atom sensors are extensively used as precision tools for detecting external fields, e.g. of magnetic and gravitational origin, which have found numerous scientific and industrial applications \cite{Alonso2022}. Atomic systems have also been deployed as sensors for highly precise clocks, significantly improving timekeeping standards and enabling advancements in navigation, telecommunication, and fundamental physics research \cite{Herrmann2018,Peil2013}. 
It is a long-standing goal to make this sensing technology available in space, both for commercial and societal applications as well as for fundamental science. 
Applications include tests of Einstein's equivalence principle~\cite{Battelier2021}, the development of atom-based gravitational wave detectors~\cite{Dimopoulos2009}, satellite-based earth observation ~\cite{Douch2018,Carraz2014,Chiow2015,Leveque2019} and navigation~\cite{Jekeli2005,Canuel2006,Cheiney2018,Barrett2019,Gersemann2020}.
Decisive advantages of space-borne atomic sensors are the extended interrogation times and the low-noise environment \cite{Abend23}. While successful tests have demonstrated the feasibility of this approach in microgravity environments \cite{Muntinga2013,Becker_2018,Condon2019,Pelluet24}, a sensor aiming for the highest achievable sensitivity must utilize squeezing and entanglement to suppress the effects of quantum noise.~\cite{Feldmann2023}.

The INTENTAS project aims to demonstrate such an entanglement-enhanced atomic sensor while meeting all space, power, and weight (SWaP) requirements for the microgravity environment provided by the Einstein-Elevator in Hannover, Germany. \\
The basic design concept for the sensor is to generate Bose-Einstein condensates (BECs) with up to 10$^6$ atoms using an all-optical approach deploying time-averaged dipole trap potentials~\cite{Roy16}. Then, the entanglement will be generated via spin changing collisions~\cite{Lucke2011,Kruse2016,Cassens2024}. In order to achieve an experimental sequence providing enough time for the spin dynamics and subsequent measurements in free fall, the system will be required to provide a 3D-MOT with 10$^9$ atoms within a second and allowing for an additional second for the BEC creation \cite{Hetzel2023}. 
In this article, we present the design of the apparatus, which is currently under construction. It serves as a reference for its subsystems and summarizes key design decisions relevant to the growing number of projects developing atom sensors for both compact terrestrial and space applications. To highlight the advantages of entanglement-enhanced sensitivity, we also provide an example performance estimation for microwave spectroscopy, which is fundamental for both microwave frequency standards and Raman-based atom interferometers.
The INTENTAS project paves the way for exploiting entanglement-enhanced interferometric sensitivities in a new generation of spaceborne atomic sensors.

\section{Physics package overview}

\begin{figure}[htbp]
    \centering
    \includegraphics[width=0.95\textwidth]{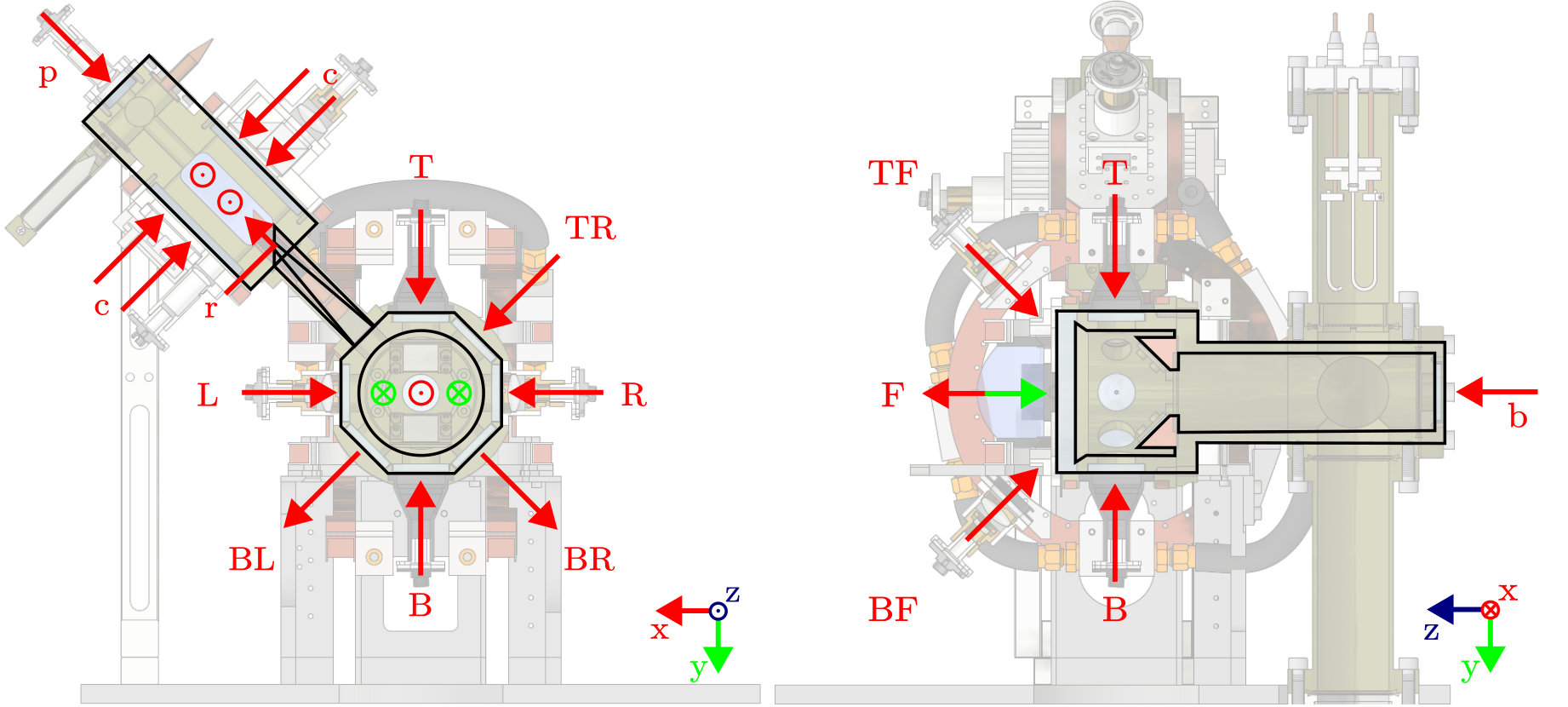}
    \caption{Sketch of the optical ports of the vacuum chamber (left: $x$-$y$ plane, right: $y$-$z$ plane): p: pusher, c: cooling, r: retarder, T: Top, TR: top right, R: right, BR: bottom right, B: bottom, BL: bottom left, L: left, TF: top front, F: front, BF: bottom front, b: back.}
    \label{fig:overview}
\end{figure}

The essential systems of the INTENTAS apparatus are shown in Fig.~\ref{fig:overview} with the orientation of the coordinate system indicated. The design is largely based on the QUANTUS and MAIUS projects which both have been successfully operated in microgravity environments \cite{Rudolph_2015,Becker_2018}. \\
The setup can be subdivided into three sections: \\1) The source chamber consists of an atom oven providing the atoms for experimentation and, additionally, a two-dimensional magneto-optical trap including a directional laser beam able to transport pre-cooled atoms into the science chamber (2D$^+$-MOT).

2) The central chamber, which is referred to as the science chamber, is where the actual atom cooling, BEC creation, and state manipulation is performed. The chamber is equipped with a total of ten ports. Seven of the ten ports have an inner diameter of 20\,mm and are equipped with windows that feature an anti-reflective (AR) coating for 780\,nm light. Two of those ports (L and R) are used for two counter-propagating laser beams enabling cooling of atoms in the $x$-direction as part of the three-dimensional magneto-optical trap (3D-MOT). Another two (TR and BL) are foreseen to be used for an absorption detection channel and two more (T and B) can be used for either fluorescence-inducing beams as a secondary detection channel or to include a Raman laser system for atom interferometry. Port BR is left open and can be equipped with additional diagnostics such as a photo diode or another auxiliary camera system. The eighth port differs from the others in that it has a CF-16 knife edge included and is used to connect the source chamber. 
The remaining two ports are larger in diameter (74\,mm and 66\,mm). Port F located in the front in Fig.~\ref{fig:overview} is equipped with a window. It is AR coated for 780 nm and for 1064 nm. Through this window two beams of the 3D-MOT enter the chamber at a 45\,\textdegree~angle (TF and BF). Mirrors inside the science chamber retro-reflect those beams right back to their origin, thus completing the three dimensional cooling of the 3D-MOT.
The front window is additionally used to inject two dipole trap laser beams (1064\,nm) and to provide optical access for the fluorescence imaging system (780\,nm). This is achieved by adding a fitted lens right at the window and a dichroic mirror separating the two light paths. \\
3) The support chamber is connected to the second large port on the opposite side of the detection lens. It houses the vacuum pumps and vacuum gauges. Furthermore, it is also equipped with a viewport (b) to enable further optical access to the science chamber. \\
The different chambers are made of titanium in order to comply with the requirements on magnetic cleanliness and the windows made of N-BK7 are connected via indium seals ensuring compactness of the setup. The source chamber and the science chamber are surrounded by various coils to create the necessary magnetic, radio-frequency (RF) and microwave (MW) fields. In case of the science chamber the magnetic coils are not only used for the MOT, but also to generate three-dimensional offset magnetic fields. A dedicated coil can generate magnetic field gradients sufficient for the separation of atoms according to their magnetic Zeeman level (Stern-Gerlach like) to enable a spin-dependent detection. The coil in the $y$-direction provides a quantization axis for Zeeman splitting. \\
RF and MW antenna are placed close to the front window to enable manipulation of atomic states.
The entire setup is surrounded by passive magnetic shielding and auxiliary systems (e.g. power supplies, amplifier) are located on the outside. The 780\,nm and 1064\,nm laser light is generated outside as well and guided to telescopes at the setup via fibers. The entire apparatus is built on a three-stage platform fitting the transport capsule in the Einstein-Elevator (EE), the support environment enabling measurements in microgravity. 

\section{Environment: Einstein-Elevator}

\begin{figure}
    \centering
    \includegraphics[width=0.95\textwidth]{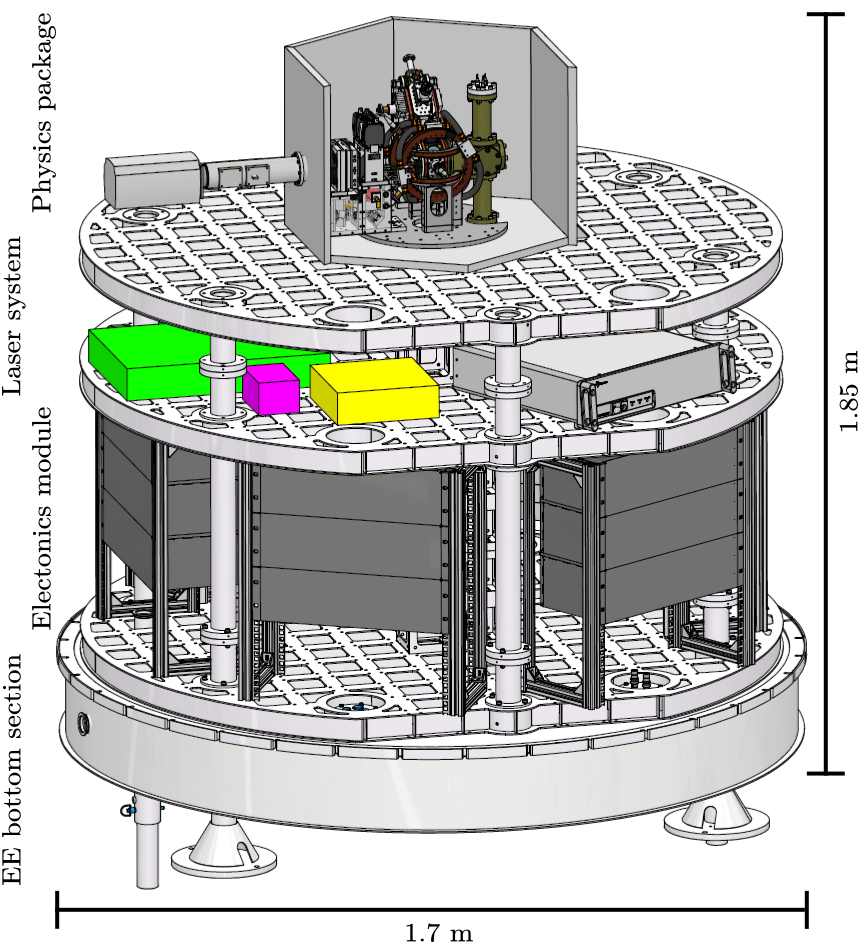}
    \caption{The INTENTAS setup with its three chambers within the magnetic shielding on top of the carrier levels, which will be installed inside the Einstein-Elevator. The racks housing the electronics and an extra level for laser systems and other critical equipment is shown. }
    \label{fig:overview_in_EE}
\end{figure}

The primary environment for which the apparatus is designed is the so-called Einstein-Elevator (EE) located at the Hannover Institute of Technology (HITec), Germany. Here, a chamber, so-called gondola, is accelerated vertically with initially about 5\,g using a precise linear drive system inside a 40\,m high tower. Before it gets decelerated with about 5\,g, it offers about 4\,s of microgravity on the level of $10^{-6}\,$g \cite{Einstein_Elevator}. The system is able to provide up to 300 launches per day, only limited by a cool-off period of about 4\,minutes. This allows for acquisition of notable amounts of data compared to classical drop towers (see Tab.~\ref{tab:merged-microgravity}). During the idle time on ground a total of 1\,kW of cooling power is provided to the water cooling system of the apparatus along with connection to the power grid to recharge batteries and run calibration measurements. Additionally, the apparatus is accessible on short time scale, which is possible since only the gondola is evacuated to about 1\,Pa instead of the entire tower. This operational concepts forsees another inner capsule (pressure hull) in order to provide an environment of atmospheric pressure for the device and all of its subsystems. This relaxes the requirements for equipment substantially. This inner capsule weighs 120\,kg and provides 167\,cm (in diameter) $\times$ 200\,cm (in height) of inner space. It is mounted to a lower section which includes electronics, sensors, control-systems and a water-pump for providing the cooling water. To run an experiment in the EE, all components are mounted on stacked carrier levels on top of an apparatus-independent bottom section. The weight of this section is 180\,kg (without included components) while a carrier level weighs about 50\,kg. Since the total weight allowed in the gondola is 1,000\,kg, and three levels are foreseen, this leaves an allocated weight of 550\,kg for the experimental setup including all auxiliary equipment. Operational parameters as the acceleration, microgravity quality, rotation and magnetic field strength can be obtained live, which enables dedicated calibration runs as well as a full monitoring of an experimental run. A CAD sketch illustrating the full experimental setup ready for the EE is shown in Fig.~\ref{fig:overview_in_EE}.

\begin{table}[h!]
\centering
\caption{Comparison of microgravity facilities sorted by various metrics. Based on \cite{Raudonis2023}.}
\label{tab:merged-microgravity}
\begin{tabular}{|l|c|c|c|}
\hline
\textbf{Facility}                 & \textbf{\textmu{}$g$ quality [$g$]} & \textbf{\textmu{}$g$ duration} & \textbf{Repetition rate} \\ \hline
Satellite                         & $< 10^{-6}$                  & days to years           & continuous                        \\ \hline
International Space Station       & $10^{-5}$                    & months to years       & continuous                        \\ \hline
Sounding rockets                  & $10^{-5}$                    & up to 13 minutes                         & continuous             \\ \hline
Parabola flights                  & $10^{-2}$          & 20 seconds                & 31 per flight      \\ \hline
Drop Tower Bremen                 & $10^{-5}$                    & 4.7 to 9.3 seconds        & 3 per day                \\ \hline
Einstein-Elevator                 & $10^{-6}$                    & 4 seconds                 & 300 per day              \\ \hline
\end{tabular}
\end{table}

\section{Auxiliary systems}

\subsection{Power system}
The power in-flight is supplied by an array of super-capacitors (HY-CAP 500F 3V) which are connected in series for each of the required voltages (24\,V, $\pm$18\,V, 12\,V, 5\,V). The system is build into two 19-inch racks. The expected voltage loss during a 4$\,s$ flight is less than 10\,$\%$, which is recharged using power supply units (R\&S HMP4040) while the system is on ground.

The voltage of each capacitor is capped at 2\,V to 2.5\,V to increase its lifespan. The resulting available energies for a given voltage and the consumption during flight are given in Tab.~\ref{tab:powersupply-energy}. To ensure proper heat dissipation, the capacitors and their circuits are cooled using fans directed towards a water-cooled aluminum lamella heat exchanger. 
\begin{table}[b!]
\centering
\caption{Total and consumption (in 4\,s) energy of the super-capacitor power supply.}
\label{tab:powersupply-energy}
\begin{tabular}{|c|c|c|}
\hline
\textbf{voltage supplied [V]}                 & \textbf{Total energy [Wh]} & \textbf{Consumption energy [Wh]} \\ \hline
24&3.3&0.146\\ \hline
$\pm$18&2.5&0.032\\ \hline
12&1.7&0.13\\ \hline
5V&0.6&0.057\\ \hline
\end{tabular}
\end{table}

\subsection{Vacuum system}

The main vacuum for the system is provided by an ion-getter-pump (VACION PLUS 20 DIODE), which achieves a pumping speed of 27\,l/s (N$_2$). It is supported by a titanium sublimation pump (ZST22L110 VACGEN) which is irregularly refreshed. Before the first operation, the system is baked uniformly at 60\,\textdegree C and evacuated with a turbo molecular pump, which is removed afterwards from the system by pinching off the copper tube connections. The pressure can be monitored by a cold-cathode gauge (Agilent IMG-300) which is located directly at the support chamber but will be turned off during science operation to avoid spurious magnetic fields. The 3D-MOT is separated from the \twoD by a differential pumping section as used in \cite{Seidel14}. This device of 63.5\,mm length includes a graphite capillary which tapers from 8.8\,mm to 1.5\,mm over its extent and thereby reduces the pumping speed such that the pressure in the \twoD chamber is about three orders of magnitude higher compared to the science chamber. Additionally, due to the graphite inserts, an enhanced absorption is expected for Rb atoms, reducing the partial pressure of \Rb in the science chamber further. The target pressure at the designated location of the atoms is below 10$^{-8}$\,Pa.

\subsection{Magnetic shielding}

\begin{figure}[htbp]
\centering
\includegraphics[width=14cm]{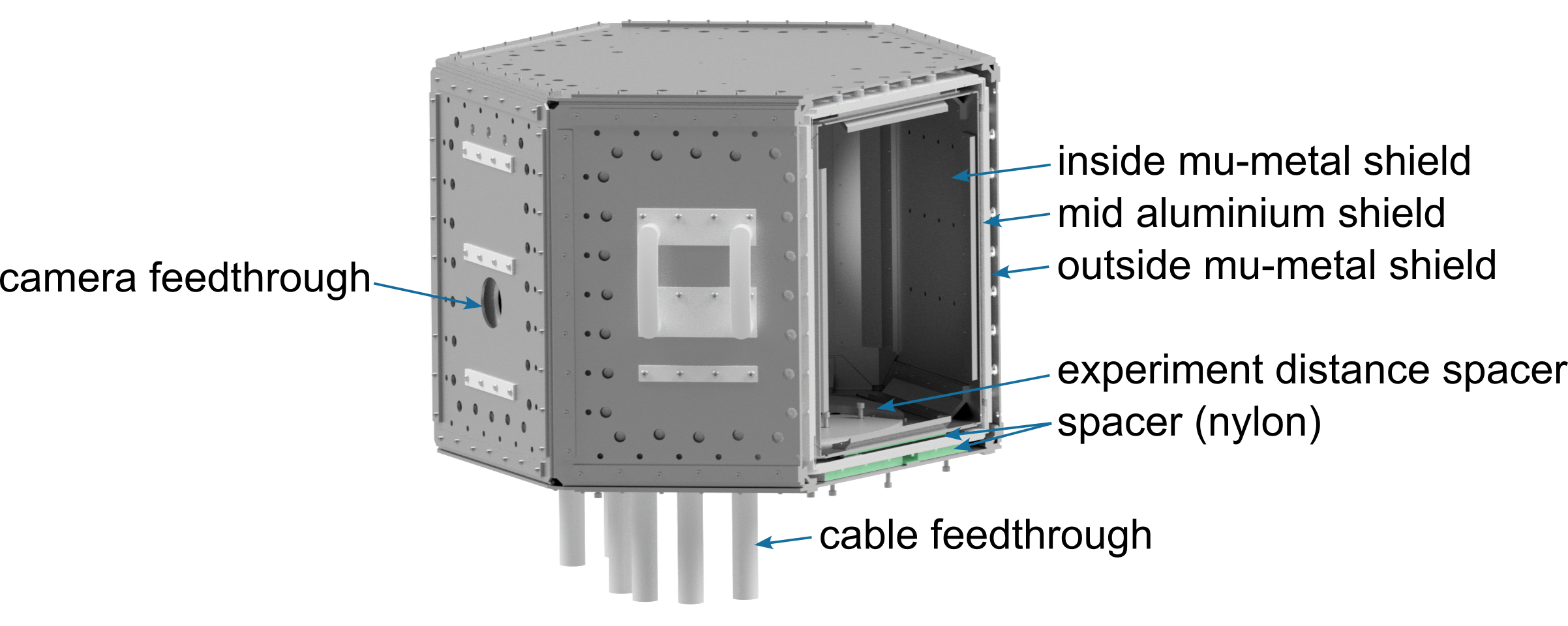}
    \caption{The CAD representation of the magnetic shield with a sectional view from a lateral perspective. Design and figure in cooperation with IMEDCO AG.}
\label{fig:Magnetic_shielding}
\end{figure}

\begin{figure}[htbp]
    \centering
    \includegraphics[width=0.95\textwidth]{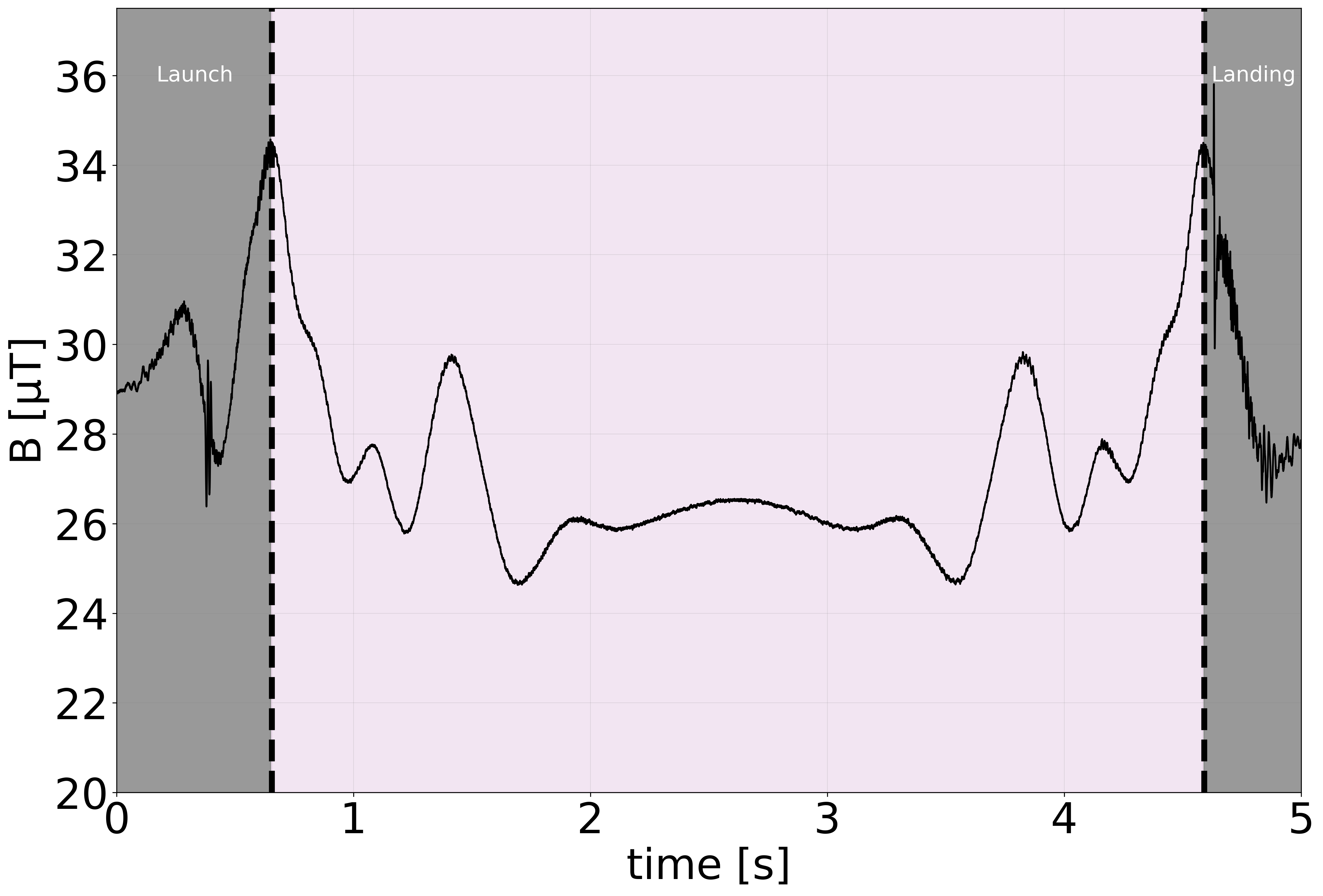}
    \caption{Measurement of the magnetic field strength at various locations sampled during a flight.}
    \label{fig:bfield_ee}
\end{figure}

\begin{figure}[htbp]
    \centering
    \includegraphics[width=0.95\textwidth]{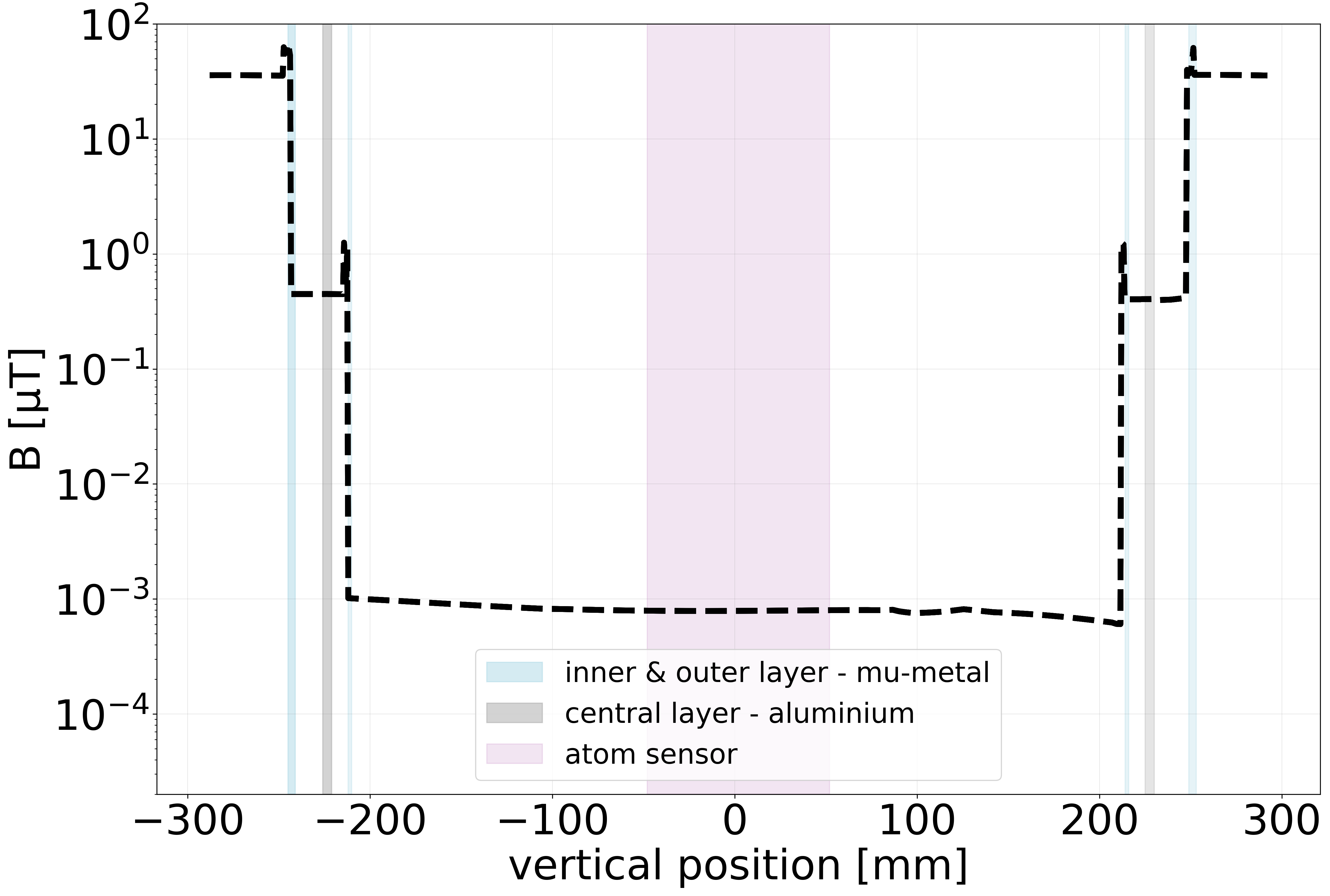}
    \caption{Simulated magnetic field strength at a frequency of $0.1\,$Hz as a function of position along the vertical axis of the experimental setup. Using the measured data (see \ref{fig:bfield_ee}), the simulation shows the remaining magnetic field inside the shielding (dashed line). The simulation reveals that the other two axes perform similar or better than the example shown here.}
    \label{fig:Magnetic_field_Ansys}
\end{figure}

In order to provide reproducible conditions in the experiment, any influence from changing magnetic fields have to be minimized. This is emphasized in an application where the external magnetic fields are changing due to the spatial movement of the experiment along different structures of the building. A passive magnetic shield is chosen to obtain an independence of the external conditions without the need of additional external coils for active compensation.\\
The magnetic shield consists of three different layers which form a hexagonal chamber closing off the system in all directions (see Fig.~\ref{fig:Magnetic_shielding}).
The inner and outer layer of the walls are made out of twelve (inside: 404\,mm\,$\times$\,298\,mm$\,\times$\,2\,mm, outside: 488\,mm\,$\times$\,364\,mm\,$\times$\,4\,mm) mu-metal plates and provide shielding for static and low-frequency magnetic fields. The center structure, which is used to anchor the mu-metal plates, is made out of aluminum of size 446\,mm\,$\times$\,327\,mm\,$\times$\,5\,mm, which provides further shielding against high-frequency (30\,Hz\,-\,100\,MHz) magnetic fields. The magnetic shielding forms a volume of 0.176\,m$^3$ and with its additional top and bottom section encapsulates the entire sensor head. In order to provide an optical access for the detection system a tube of length 15\,cm is added. In addition, six feedthroughs with the length of 19\,cm were placed for optical fibers, cables and cooling lines, as well as a dedicated feedthrough for the ion-getter pump, which also acts as electrical ground for the science chamber. The mu-metal plates will be annealed before usage and equilibrated in all three dimensions using copper coils in an L-shape configuration around the inner and outer walls of the mu-metal in order to minimize build-up of residual magnetic fields \cite{I_Altarev_P_Fierlinger_2015}. 
The performance of the shield design has been simulated using ANSYS 2022. Therefore, the external magnetic field in the Einstein-Elevator has been measured and evaluated in a test flight using a magnetometer (Bartington Mag-03) as shown in Fig.~\ref{fig:bfield_ee}. The evaluation has revealed that the magnetic field strength change is less than $\pm$5\,\textmu{}T along the entire trajectory, while the largest offset field is around 40\,\textmu{}T. Using those values, we find that the presented design of the magnetic shield leads to a substantial reduction for the measured magnetic field (see Fig.~\ref{fig:Magnetic_field_Ansys}). Within the shield, at the location of the atoms, the mean magnetic field strength according to the simulation is expected to be reduced by a factor $>10.000$ for low-frequency fields (0.1\,Hz) and is suppressed further with increasing frequency. This will yield a magnetic field strength of $<10\,$nT at the location of the atoms while operating in the Einstein-Elevator.

\section{Atom cooling and trapping}

Cooling of $^{87}$Rb is performed by slightly off-resonant 780\,nm laser light provided by a dedicated laser system designed to meet the special needs of the EE. Trapping of the atoms is achieved by a combination of applied magnetic fields and laser light initially and finally utilizing an optical dipole trap. The optical dipole trap increases the phase space density necessary for BEC creation by taking advantage of time-averaged potentials. 

\subsection{Laser system}

\begin{figure}[htbp]
\includegraphics[width=0.95\textwidth]{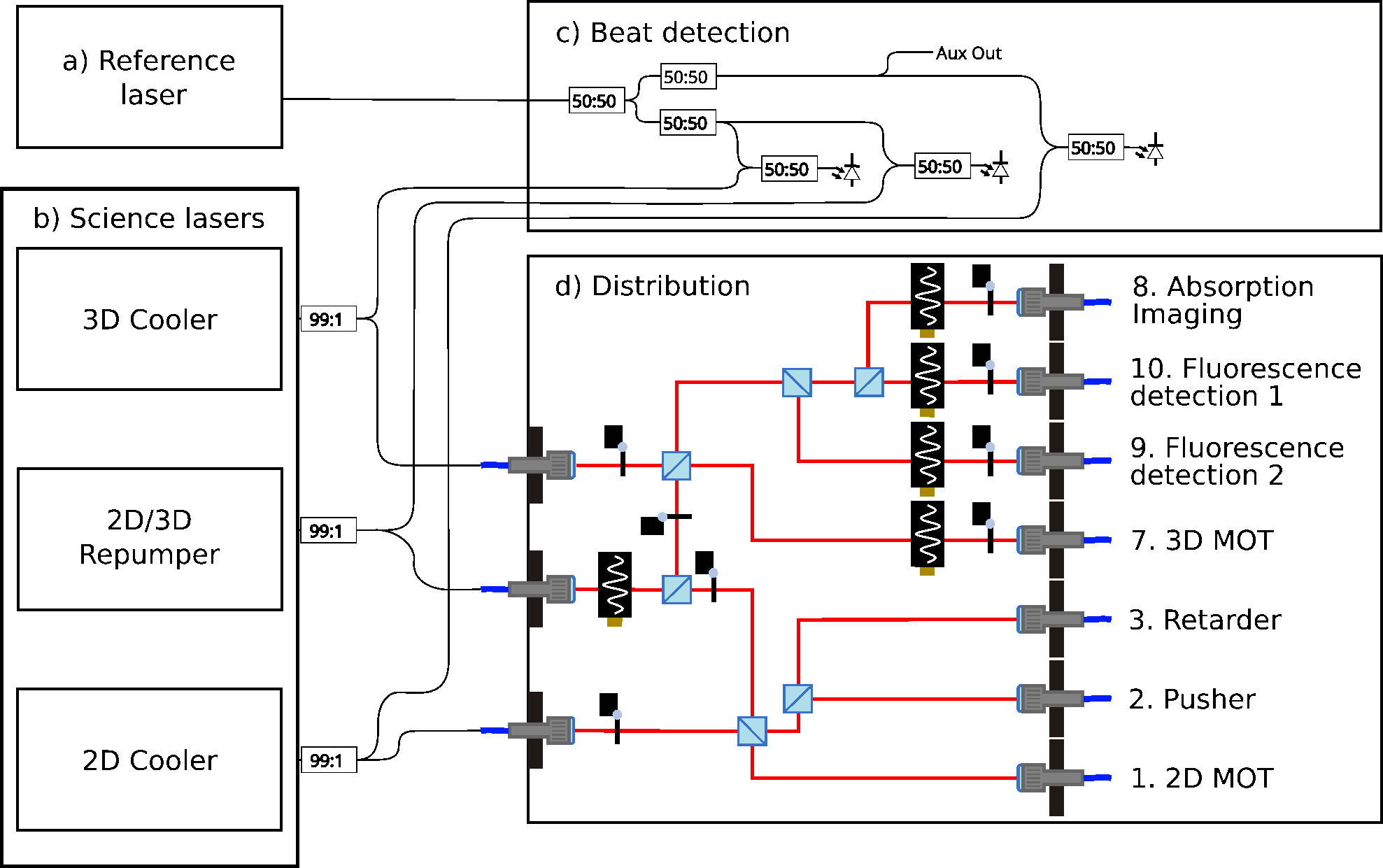}
    \caption{Schematic of the laser system. The system consists out of four subsystems: a) reference laser, b) science lasers, c) beat detection and d) the distribution system. Black lines mark connections using optical fibers and red lines mark free-space beam paths. The numbering identifies the fiber inputs and outputs as shown in Fig.\ref{fig:Distribution_System}.}
\label{fig:Schematic_Lasersystem}
\end{figure}
The EE poses a challenging environment for optical systems. Contrary to the controlled environment of laboratories, the system is subjected to accelerations, temperature drifts, vibrations as well as limited resources in terms of size, weight and power. 
To address these challenges, the laser system developed for the INTENTAS project is a combination of free-space optics and fiber-based parts. In Fig.~\ref{fig:Schematic_Lasersystem} the general optical layout can be seen. The system is divided into four subsystems: science laser sources, reference laser, beat detection and distribution system. The science laser sources are connected via optical fiber splitters with a splitting ratio of 99:1. While the major part of the light is sent to the distribution system,  1\,$\%$ of the light is directed to the beat detection system. The reference laser is directly connected to the beat detection system.

\subsubsection{Laser sources}
\begin{figure}[htbp]
    \centering
    \includegraphics[width=0.95\textwidth]{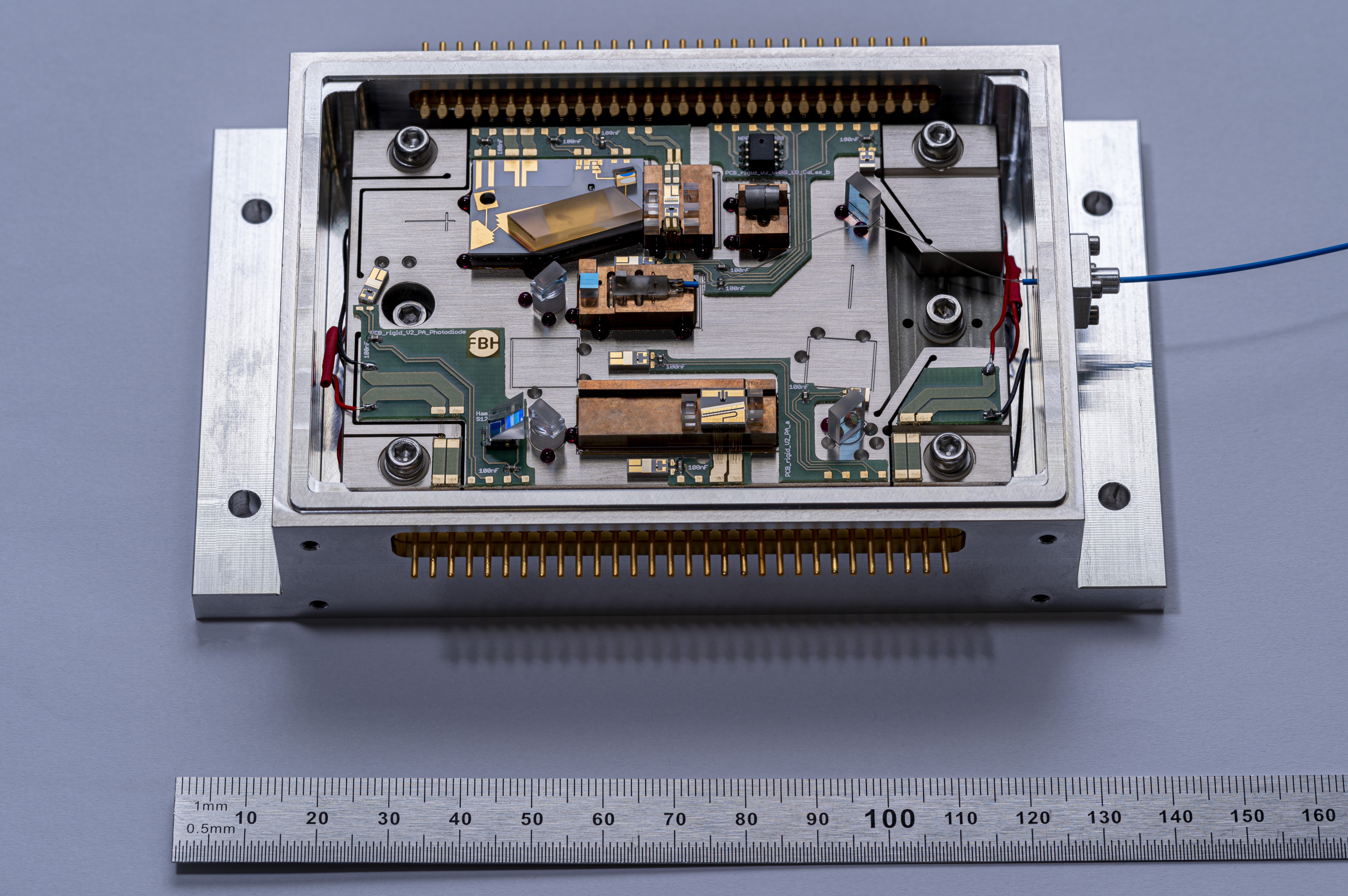}
    \caption{Photograph of a science laser module.}
    \label{fig:laser_module_photograph}
\end{figure}

In order to cope with the requirements of the EE, the science lasers are implemented as fiber-coupled micro-integrated diode laser modules using a technology based on the one described in \cite{kurbis_extended_2020,Hirsch2024}. This enables robust science laser modules (see Fig.~\ref{fig:laser_module_photograph}) with a footprint of 139\,mm\,$\times$\,80\,mm and a mass of approximately 0.8\,kg.

\begin{figure}[htbp]
    \centering
    \includegraphics[width=0.95\textwidth]{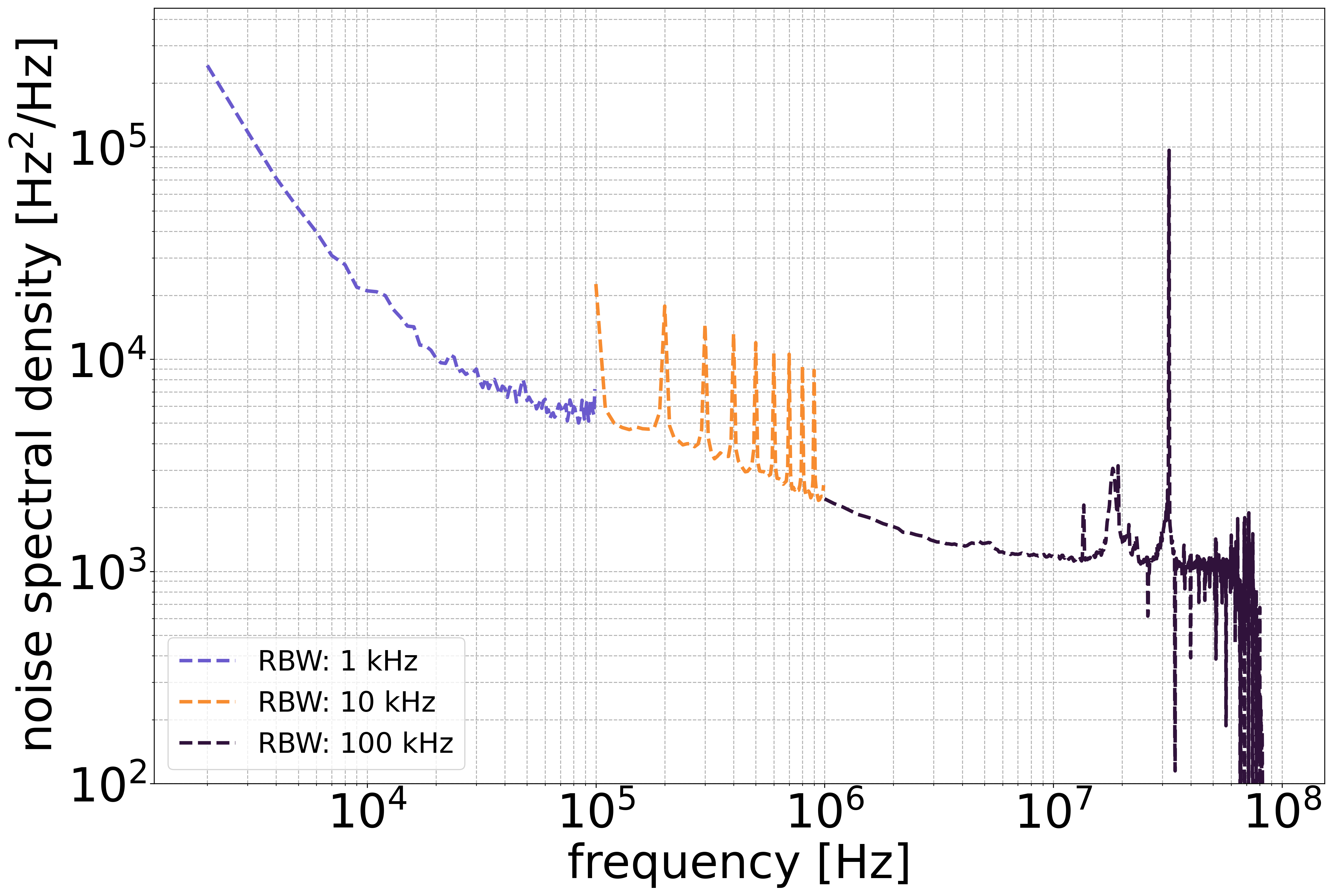}
    \caption{Frequency noise of a science laser at the working point using a self-heterodyne measurement. The peaks at multiples of 100\,kHz are measurement artifacts.}
    \label{fig:laser_fn}
\end{figure}

Conceptually, the lasers are based on a master-oscillator-power-amplifier setup. In order to achieve a low frequency noise, an extended cavity diode laser (ECDL) serves as master oscillator. The operating parameters such as the injection current applied to the ECDL ({I$_{\text{ECDL}} = 114.5$\,mA}), the temperature set points of the micro-optical bench ({T$_{\text{MIOB}} = 24$\,\textdegree C}) and the volume holographic Bragg grating ({T$_{\text{BG}} = 16.5$\,\textdegree C}) are selected so that the laser emits in a stable single mode at the desired frequency (working point). Fig.~\ref{fig:laser_fn} depicts the frequency noise of a science laser at the intended working point measured with a self-heterodyne setup. The white noise floor of the measurement \cite{schiemangk_accurate_2014} corresponds to an intrinsic linewidth of 2\,kHz. Using the beta-separation line method \cite{di_domenico_simple_2010} a FWHM-linewidth of 60\,kHz (1\,ms) can be derived. 

\begin{figure}[htbp]
    \centering
    \includegraphics[width=0.95\textwidth]{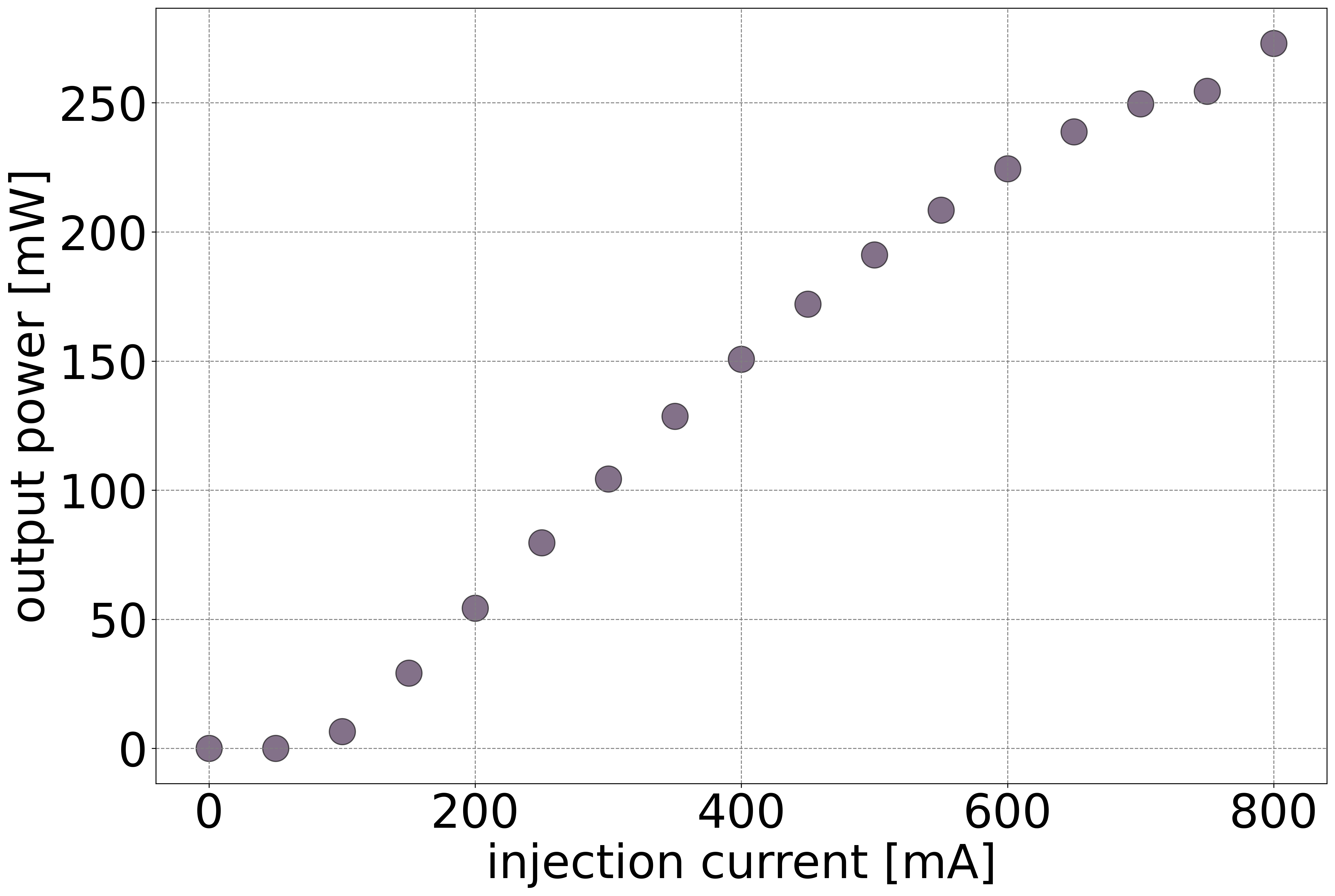}
    \caption{Fiber-coupled optical output power of a science laser vs. injection current into the ridge-waveguide amplifier chip.}
    \label{fig:laser_power}
\end{figure}

After passing through a semi-double stage optical isolator, the emission of the ECDL is amplified in a semiconductor ridge-waveguide amplifier before it is coupled to a single-mode polarization maintaining optical fiber. Fig.~\ref{fig:laser_power} shows the optical power emitted from the output fiber for varying injection current into the amplifier at the intended working point of the laser. For operating the sensor in the EE an optical output power of 250\,mW is required. Thus, an injection current into the ridge-waveguide amplifier of 750\,mA was chosen as working point.

\subsubsection{Reference laser}
\begin{figure}[htbp]
    \centering
    \includegraphics[width=0.95\textwidth]{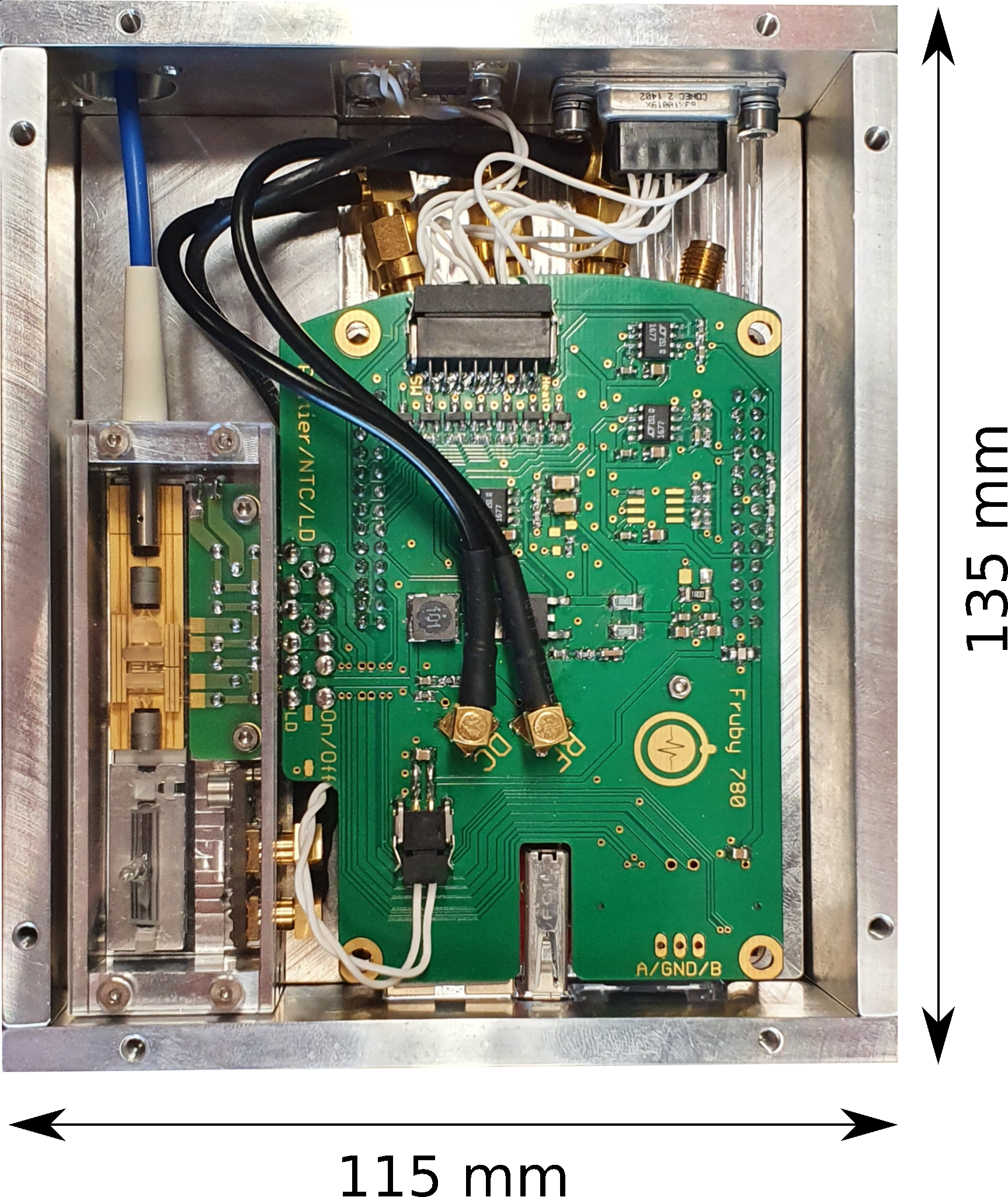}
    \caption{Photograph of the reference laser module with the micro-integrated bench containing the optical setup (left) and driver and control electronics (right).}
    \label{fig:Ref_pic}
\end{figure}

\begin{figure}[htbp]
    \centering
    \includegraphics[width=0.95\textwidth]{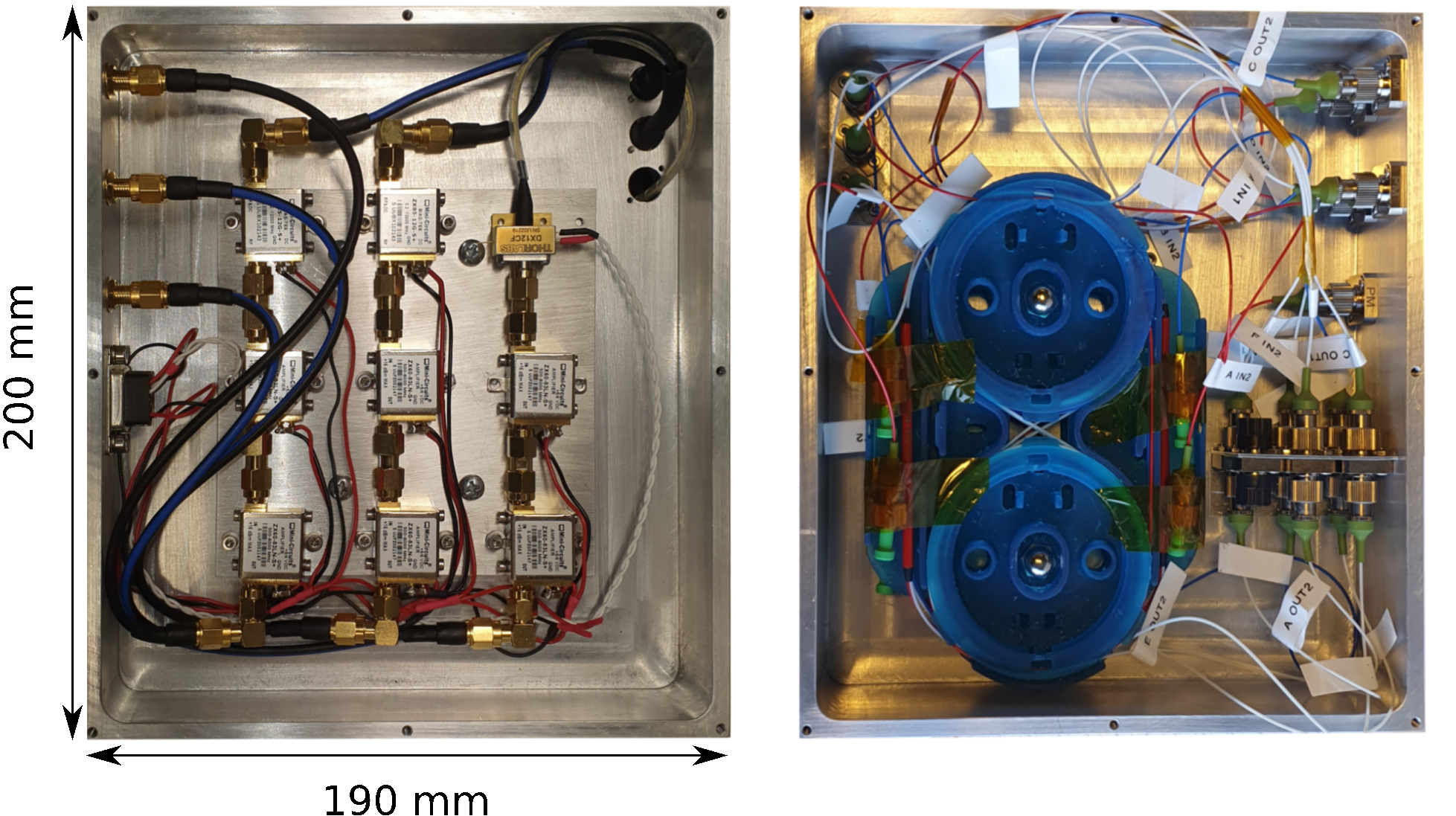}
    \caption{Photograph of the beat detection system with electronic compartment (left) and fiber splitting network in the optical compartment (right).}
    \label{fig:Beat_electronic}
\end{figure}

As reference for the science lasers, a micro-integrated frequency reference module is used. It consists of a distributed feedback laser (DFB) with a dedicated frequency modulation spectroscopy (FMS) unit based on the design in \cite{Strangfeld:21,Strangfeld:22}. The design has been adapted to the needs of the sensor by modifying the laser and spectroscopy units and the included driver and control electronics. 
The module provides 15\,mW of optical power via a polarization maintaining fiber and comes with a built-in laser current driver, temperature controller as well as lock electronics. A photograph of this subsystem can be seen in Fig.~\ref{fig:Ref_pic}. 
The module is operated via a modified version of the \emph{Linien} software \cite{Wiegand2022}. It achieves a frequency stability of 1.7\,$\times$\,10$^{-12}$ after 1\,s of integration time and 1.7$\times$10$^{-11}$ at a timescale of 8\,h which was already shown in previous versions of the reference module \cite{Strangfeld:21,Strangfeld:22}. The stability of the module was measured over a time period of 13\,h, by measuring the beat note with another lab based frequency reference. The module has a size of 135\,mm\,$\times$\,165\,mm\,$\times$\,40\,mm and a mass of 3.3\,kg.

\subsubsection{Beat detection system}
The beat detection system is entirely fiber-based. The compartment housing this fiber network, is shown in Fig.~\ref{fig:Beat_electronic}. The fiber-based approach, with higher losses but also less need for maintenance, is preferred for long continuous measurement campaigns due to its expected stability gain in comparison with free-space optics. In the system, the light of the reference laser is split into four ports via an array of fiber splitters. The outputs of these four ports are mixed with the light of the science lasers in 50:50 fiber splitters. This generates the beat notes which are detected on fiber coupled photodiodes. The electrical signal is fed into the separated electronics compartment. The compartments are separated to reduce effects due to heating caused by the active electronic components. The module has a size of 210\,mm\,$\times$\,200\,mm$\,\times$\,87\,mm and a mass of 1\,kg.

\subsubsection{Distribution System}
\begin{figure}[htbp]
\centering
\includegraphics[width=0.95\textwidth]{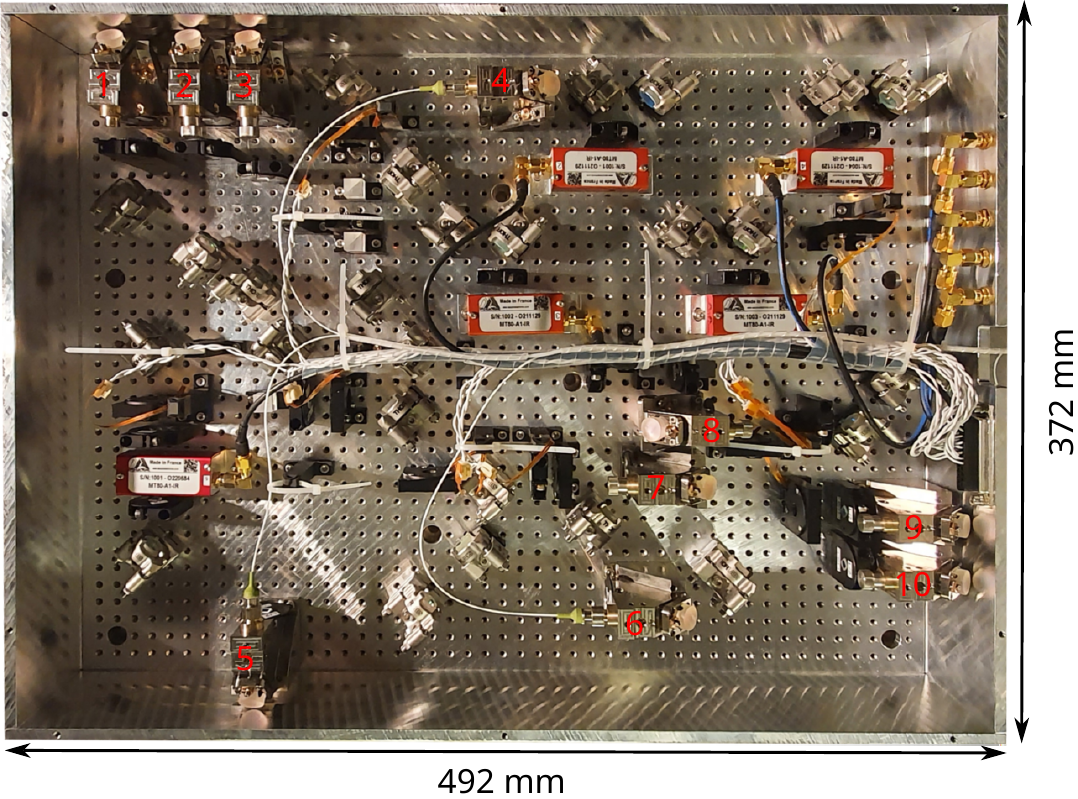}
    \caption{Photograph of the distribution system used for the INTENTAS flight system. Optical fibers are connected to the subsystem via 10 periscopes. The distribution system has a footprint of 492\,mm\,$\times$\,372\,mm$\,\times$\,95\,mm and it has a mass of 17.1\,kg. The output and input ports are labeled in Tab.~\ref{tab:List_port_distribution}}
\label{fig:Distribution_System}
\end{figure}

\begin{table}[h!]
\centering
\caption{List of input and output ports in the distribution system.}
\label{tab:List_port_distribution}
\begin{tabular}{|c|c||c|c|}
\hline
\textbf{ Number }                 & \textbf{Port} & \textbf{ Number}& \textbf{ Port} \\ \hline
1&Output \twoD&6&Input 3D-MOT cooler \\ \hline
2&Output pusher&7&Output 3D-MOT \\ \hline
3&Output retarder&8&Output absorption imaging \\ \hline
4&Input \twoD cooler&9&Output fluorescence detection 2 \\ \hline
5&Input 2D/3D repumper&10&Output fluorescence detection 1 \\ \hline
\end{tabular}
\end{table}
The major part of the output power of the  science lasers is sent to the distribution module. To achieve the high efficiencies required to provide sufficient optical power to the sensor head, a free-space approach is chosen. A picture of the system can be found in Fig.~\ref{fig:Distribution_System}. The design follows similar system designs, which already proved their reliability in the drop tower in Bremen~\cite{pahl2019,Rudolph_2015}.

The light is coupled into the distribution module using periscopes. As a next step, the 2D/3D repumper light is mixed with the \twoD cooler and 3D cooler light, respectively. The light is then distributed into different optical paths using a combination of $\lambda/2$-plates and polarizing beam splitter cubes. In each path shutters (Uniblitz type ES6B1ZM0) are placed to block the light. These shutters provide transfer times of ~2\,ms. At the end, the light is coupled into seven optical fibers using periscopes.

In detail, the light of the \twoD cooler laser is fed into the system via periscope 4 and the 2D/3D repumper laser via periscope 5. Periscope 1 is used for the \twoD beams, which is followed by a fixed 50:50 fiber-splitter, periscope 2 for the pusher and periscope 3 for the retarder. 

The light of the 3D-MOT cooling laser enters the distribution system through port 6 and is mixed with the 2D/3D repumper as well. It is used for trapping the atoms in the 3D-MOT as well as for imaging them. Each path includes an AOM for power modulation and fast switching. The deployed AOMs (AA Opto-Electronics type AA.MT80-A1-IR) provide diffraction efficiencies of 85\,\%. In combination with a single AOM at the input of the repumper laser, the power of cooler and repumper in the fiber can be adjusted independently. Periscope 7 provides light for trapping atoms in the 3D-MOT. It is followed by a 1\,$\times$\,4 fiber-splitter which equally distributes the light to the 3D-MOT beams. The periscopes 9 and 10 provide light for fluorescence detection, and periscope 8 provides light for the absorption imaging. 

The coupling efficiencies of the system are ~85-90\,\% with a PER of $<$~25\,dB. The power stability was measured to be below 0.4\,\%/s at a timescale of 10\,s. This measurement assumed the worst possible setting by using the longest optical path on the distribution board and setting all polarizing beam splitters to 50:50 splitting ratio, thus converting any polarization fluctuation into the maximal possible power fluctuation. 

\subsubsection{Ground laser system}
In order to test capabilities on the ground another fiber-coupled laser system has been developed. While larger in footprint and not optimized for usage in the Einstein-Elevator, it offers more flexibility in terms of power and modifications for the initial testing on the ground. The laser system provides 485\,mW of total power to the \twoD, where 71\,mW are split off to be available for the pusher beam and 40\,mW available for the retarder beam. For the 3D-MOT it supplies a total amount of 494\,mW which is evenly split into four paths using the same 25:25:25:25 splitter as foreseen for the final laser system (Evanescent 1\,$\times$\,4~PM 954P coupler array, Nufern PM780-HP). The two paths for the \twoD and 3D-MOT are independent of each other and share only the seed laser and an additional EOM to add a side-band for repumping during the MOT cooling stages. The 3D-MOT path also provides capabilities for the fluorescence and absorption detection. Each path is connected via an AOM and can be stabilized and shut off separately. The system is connected to a rubidium vapor cell in a modulation transfer spectroscopy module on the \twoD path. An offset frequency stabilization reaches a stability of 80\,kHz in 10\,ms, as well as tuning capabilities via an EOM to achieve the desired frequency shifts for all experimental sequence steps. It provides the capability to adjust the frequency up to 9.2\,MHz/ms. The polarization stability is 0.15° in the ellipticity and 0.32° in the azimuth over 14 hours for the 3D-MOT path. The laser system is housed in one 19-inch rack (483\,mm\,$\times$\,266 mm\,$\times$\,326\,mm) occupying 6U and is connected to the water cooling.  

\subsection{Magneto-optical trap}
\subsubsection{Optical system}

The light from the laser system is guided to the science and source chamber by polarization maintaining fibers. Self-built telescopes produce collimated beams with circular or linear polarization.
In the \twoD, the pusher beam is propagating towards the center of the differential pumping stage, pushing the atoms towards the 3D-MOT. The retarder beam counter-propagates by reflecting on the angled copper end-face of the differential pumping stage. The pusher and retarder beams have a 3.3\,mm waist and linear polarization.
The radial cooling light is coupled in perpendicular to the pusher light. Two orthogonal cooling telescopes produce a circular-polarized, collimated beam with a waist of 8.1\,mm. Each beam passes through a prism, splitting it into two parallel beams, to double the area where the atoms are exposed to the cooling light. The cooling beams are retro-reflected by two 20\,mm\,$\times$40\,mm $\lambda$/4 plates, which feature an HR coating to perform a polarization flip.

In the science chamber, four cooling beams are used to create the 3D-MOT. The telescopes are identical in construction to the ones of the \twoD cooling beams. In horizontal $x$-direction, two telescopes produce counter-propagating beams with opposite circular polarization. In the $z$-$y$ plane, two telescopes produce circular-polarized beams, that are retro-reflected within the science chamber by HR-coated $\lambda$/4 plates (20\,mm\,$\times$20\,mm).

Two detection systems are available, one for fluorescence detection and one for absorption detection. The fluorescence light is detected through the large front window of the science chamber with a lensing system described in chapter \ref{chap:detection}. To illuminate the atoms, two fiber-couplers (S+K 60FC-4-M18-10) generate counter-propagating beams along the vertical $y$-axis. The linear polarized, collimated beams have a waist of 2.9\,mm. The intensity of each beam is stabilized individually. For the absorption detection a single fiber-coupler (Thorlabs TC06APC-780) produces a single beam in the $x$-$y$ plane. The collimated linear-polarized beam has a waist of 1.3\,mm. The intensity of the absorption beam is also actively stabilized.

\subsubsection{Magnetic field coils}

Next to the optical capabilities, there are several coils added to provide magneto-optical trapping and manipulation. The \twoD is surrounded by four coils. Those are rectangular shaped, have a size of 47\,mm\,$\times$\,83\,mm, and are separated by 62\,mm. Each coil comprises 64 windings of Kapton-insulated copper wire with a diameter of 0.75\,mm.
The perpendicular coils are connected via race track configuration generating a quadrupole trap. The coils provide a magnetic field of about 56.5\,\textmu{}T/A/cm in the trap center. 
The science chamber is surrounded by a pair of coils in all three directions to compensate static magnetic fields. Each pair is set up in a Helmholtz configuration creating a homogeneous field along each axis. The coils in the $x$-direction have each a radius of 81\,mm and 19 windings and can generate a magnetic field of 0.21\,mT/A. The orthogonal coils in the $z$-direction have a radius of 49\,mm and 28 windings and can generate a magnetic field of 0.52\,mT/A. The coils in the vertical $y$-direction have a radius of 65\,mm and 12 windings, enabling a magnetic field of 0.16\,mT/A. 
Using the same support structure, there are additional coils in the $x$ and $y$-direction. The main coils creating the quadrupole field for the 3D-MOT are added in $x$-direction. They have 122 windings and are connected in anti-Helmholtz configuration. They are able to provide a magnetic field gradient of 200 \textmu{}T/A/cm in the axial direction. On the $y$ coils two additional pairs of coils with 12 and 84 windings are added. The former coils are operated in Helmholtz configuration in order to generate a stable quantization field. The latter ones provide a short burst in order to separate the atoms according to their different m$_F$ states. The achievable gradient in anti-Helmholtz configuration is 215 \textmu{}T/A/cm. The thermally relevant coils at the science chamber are connected to a water cooling system via cooling pads attached to the coil support structure.

\begin{table}[h!]
\centering
\caption{Summary of all coils used for magnetic fields.}

\begin{tabular}{|l|c|p{2cm}|p{2.5cm}|c|}
\hline
\textbf{Coil} & \textbf{Windings} & \textbf{Size (mm)} & \textbf{Purpose} & \textbf{Field} \\ \hline
2D & 64 & 47x83 (rect.) & Quadrupole field & 56.5 \textmu{}T/A/cm \\ \hline
$x$ Helmholtz & 19 & 81 (radius) & Compensation field & 0.21 mT/A \\ \hline
$z$ Helmholtz & 28 & 49 (radius) & Compensation field & 0.52 mT/A \\ \hline
$y$ Helmholtz & 12 & 65 (radius) & Compensation field & 0.16 mT/A \\ \hline
3D-MOT & 122 & 81 (radius) & Quadrupole field & 200 \textmu{}T/A/cm (grad) \\ \hline
Y Quantization & 12 & 65 (radius) & Quantization field & 0.16 mT/A \\ \hline
Y Stern-Gerlach & 84 & 65 (radius) & Separate mF states & 215 \textmu{}T/A/cm (grad) \\ \hline
\end{tabular}

\label{table:coils}
\end{table}

The currents through the coils are controlled by a self-built current driver based on \cite{Popp_2018}. The newest version is compatible with the ARTIQ Kasli FPGA and can therefore be directly controlled by the experimental control (see \ref{chap:exp_control}). The current drivers deliver a current of $\pm$10\,A at a noise of 2$\times$10$^{-6}$\,$A_{\text{RMS}}$/${\sqrt{\text{Hz}}}$ in a frequency range between 1\,Hz and 12.5\,kHz and simulation shows that the currents are reached within a few ms.

\subsection{Dipole trap}

The cODT is powered by two fiber-based 1064\,nm lasers (NKT Basik + NKT Boostik HP) delivering a power of up to 15\,W per beam. Each generated beam is expanded to a beam radius of 1.95\,mm (1/ $e^2$) by a lens system (Thorlabs LC1054-C and Thorlabs AC127-075-C) and then reflected by two mirrors into an acousto-optical-deflector (AOD) (AAOptoelectronics DTSXY-400). These have an efficiency of $>$50\,$\%$, a maximum input power density of 10\,W/mm$^2$, a center frequency of 75\,MHz, and a frequency range of $\pm$15\,MHz. The resulting beam can be deflected up to an angle of $\pm$\,1.4\,\textdegree~by changing the RF frequency operating the AOD. Both beams are then focused by the same aspheric lens (Asphericon ALL75-60-P-U-780) to the intersection point and cross under an angle of 30\,\textdegree. The resulting focal point has a diameter of 10.5\,\textmu{}m (1/$e^2$) and a Rayleigh range of 320\,µm. With this system it is possible to bring up to 10\,W per beam to the location of the atoms. 

To increase the trapping volume of the cODT, both beams can be spatially modulated in two directions. Given the capabilities of the used AODs, a stroke of $\pm$1.4\,mm perpendicular to the beam is possible. The modulation frequency to create a time averaged potential is provided by an RF signal sourced from a software-defined radio (Ettus USRP X410), which allows to configure various trap geometries \cite{Hetzel2023,Herbst2024}.\\
The power of the light going to the atoms is actively stabilized. Therefore a fraction of the light is collected on a photo diode (Thorlabs PDA100A2). The photo diode voltage is used as signal for a PID controller, the analog set point is given by a DAC (Sinara Zotino). The output of the PID is used as control voltage for a voltage-controlled attenuator (Minicircuits ZX73-2500-S+) which can attenuate the RF signal from the RF source and therefore regulates the laser power. To fully switch off the trap, additional RF switches (Minicircuits ZYSWA-2-50DR+) are implemented. The system is connected sturdily with the science chamber and has a custom made aluminum housing to minimize relative drifts of the two dipole trap beam paths. This has been verified in a test flight in the EE. \\
In addition to the function as an initial trap the dipole laser system offers also the possibility to apply delta-kick collimation to the atomic ensembles during the detection phase \cite{Ammann97}. 

\subsection{Detection system}
\label{chap:detection}

The setup has two detection schemes to achieve full spatial resolution. The fluorescence detection is recorded by a CCD camera with a quantum efficiency of 95\,$\%$ at 780\,nm (Andor iKon-M934). The lens system imaging the atoms on the chip consists of four lenses. The first one is the same lens that focuses down the dipole trap beams. The detection is separated after this lens from the dipole trap beam path by a di-chroic mirror (Focktec DCF7575-780-1064). The second and the third lens are aspheric lenses (Thorlabs AL75150H-B and AL50100H-B). The fourth lens is custom manufactured (50\,mm in diameter with AR-coating for 780 nm, made out of H-K9L). This is necessary to compensate the influence of the vacuum window. The magnification factor of this system is simulated to be 1.7. The resolution therefore should be limited to 7.7\,\textmu{}m by the pixel size of 13\,\textmu{}m.\\
A second detection channel for an absorption detection scheme, which is oriented orthogonal to the fluorescence detection from the top right to the bottom left window and can be added.

\section{State manipulation}

The experimental setup provides atoms that can be transferred into different spin states. This allows for applications like the entanglement generation in BECs via spin changing collisions. Hereby RF and MW pulses are necessary in order to enable the desired state transfer. MWs can be used for the initial preparation into specific hyperfine levels, the dressing of Zeeman states or for the manipulation of the states in an interferometric sequence as well as for the rotation and detection of specific states. 
\subsection{Microwave source }

\begin{figure}[htbp]
	\centering
    \includegraphics[width=0.95\columnwidth]{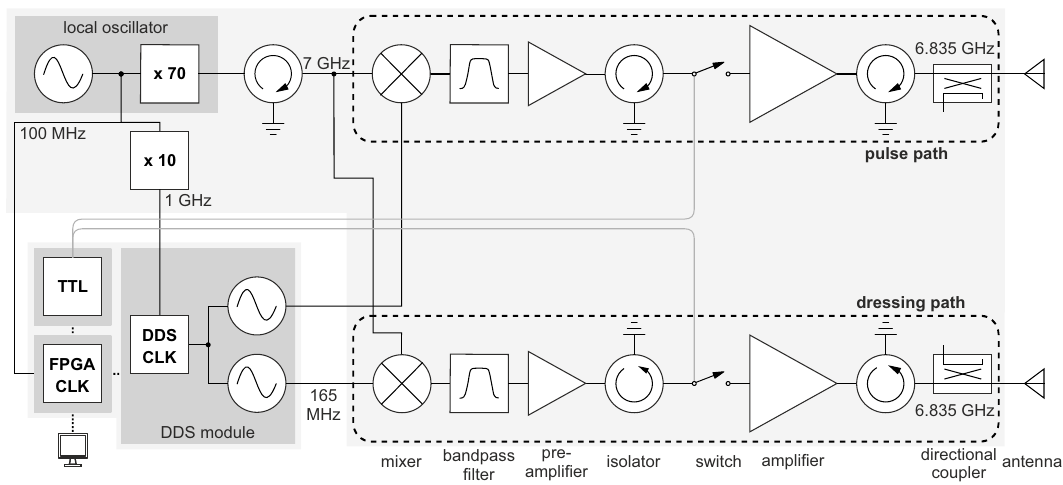}
    \caption{Schematic overview of the microwave source. Modified with permission from \cite{meyer-hoppe23}.}
	\label{fig:Schematic_MW}
\end{figure}

The source providing the MW used for dressing and transitioning magnetic hyperfine states is a copy of the system presented in \cite{meyer-hoppe23}. The underlying schematic of the system is shown in Fig.~\ref{fig:Schematic_MW}. It provides two independently controllable output paths. Each path is able to provide 6.835\,GHz with a low integrated phase noise of 480\,\textmu{}rad between 10\,Hz and 100\,kHz and an upper bound on the relative amplitude noise of 0.03$\%$. The system provides tunability in order to address different states depending on magnetic field of about 25\,MHz. Additionally, it allows for fast dynamical adjustments in less than a \textmu{}s. All of this is possible by combining a low phase noise oscillator with a configurable FPGA. 

Two microwave antennas are placed perpendicular to each other. This way circular and linear polarized microwaves can be applied to the atoms. Around each of the telescopes of the retro-reflected 45°-MOT beams a single winding of copper wire with about 7.6\,mm radius is placed. Their circumference is matched to the 6.8\,GHz transition between the F=1 and F=2 level of the $^{5}$S$_{1/2}$ level of $^{87}$Rb.

\subsection{RF antenna}

The system has three RF antennas. Two of them are perpendicular to be able to generate RF signals with circular polarization. They have a radius of 3.5\,mm, a winding number of 2 and a distance of 51\,mm to the atoms. The third RF antenna has a radius of 43.5\,mm, a winding number of 1 and a distance to the atoms of 29\,mm. 

The RF signal is supplied by a DDS and then amplified by a power amplifier (AA opto-electronic: AMPB-B-34-10.500). The required frequencies are in the range of the Zeeman splitting of \Rb in a quantization field of 10\,\textmu{}T which are around 700\,kHz. 

\section{Experimental operation }

The experimental sequence needs precise timing and control of frequencies, intensities, currents and voltages. Furthermore, the operation in the EE requires a fully remote control and monitoring system.  

\subsection{Experimental control}
\label{chap:exp_control}
The control system is designed towards usage of the Advanced Real-Time Infrastructure for Quantum physics (ARTIQ) system which includes the Sinara hardware environment \cite{Artiq2021}\cite{sinara24}. The hardware components of this system are variable and are only limited by spatial requirements of the capsule in the elevator. In general there is the possibility to use up to 3 FPGA modules (`Kasli'), which allow several input/output operations via sub modules (e.g. DAC module `Zotino'). Furthermore, various RF frequency ramps of a few hundred MHz can be achieved using dedicated DDS devices (`Urukul' \cite{Kasprowicz2022}) and higher frequencies (at least up to 12 GHz) are generated by the microwave synthesizer (`Mirny'), which can be updated on the \textmu{}s scale as well. Another module (`Stabilizer') acts independently as a PID control for the frequency stabilization and intensity stabilization of the lasers. All of this will be mounted into 19-inch rack spaces with custom made water cooling boards to dissipate the produced heat efficiently in the encapsulated environment.

Next to the real-time system controlling the experimental sequences, there is a monitoring system based on the LabJack T7-Pro. This so-called slow control system is able to monitor relevant environmental parameters, e.g. temperature, and gives control over various hardware devices. The user interface for this relies on the Grafana environment \cite{Grafana2019}.

\subsection{Experimental sequence}

\begin{figure}[htbp]
	\centering
    \includegraphics[width=0.95\columnwidth]{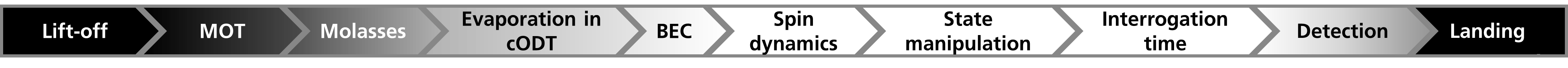}
    \caption{Schematic overview of the experimental sequence. }
	\label{fig:Schematic_ES}
\end{figure}

The experimental sequence (see Fig.\,\ref{fig:Schematic_ES}) starts with loading pre-cooled atoms from the \twoD into the science chamber, where the
3D-MOT is filled. The source of atoms is the atom oven, which can optionally be heated if the amount of residual Rb-atoms in the \twoD is not sufficient. The \twoD is started to generate a pre-cooled atom beam towards the 3D-MOT guided by the pusher beam. The velocity of the atoms can be adjusted by the counter-propagating retarder beam. Simultaneously, the 3D-MOT is turned on in order to fill the trap region with the incoming atoms. A loading rate of 1$\times$10$^{9}$\,atoms/s should be achievable. Once the trap is sufficiently loaded the magnetic fields of the 3D-MOT and \twoD as well as the laser light for the latter are turned off. The light of the 3D-MOT is then gradually red detuned and reduced in intensity to generate an optical molasses, which should cool down the atoms towards a temperature less than 50\,\textmu{}K in all three dimensions.
In parallel, the laser beams of the cODT can be turned on and initialized with the largest painting stroke assuring the highest possible occupation in the cODT. Once the molasses and repump beam are turned off, the painting stroke and intensity of the cODT beams is reduced simultaneously to induce evaporative cooling. It can be assisted further by providing a magnetic field to selectively remove hot atoms from the magnetic sensitive states (so called spin-distillation cooling). This process should take about one second and the final phase space density should be sufficient to allow for BEC creation. Afterwards, MW and/or RF pulses can be used to manipulate the states and induce spin dynamics via dressing either in the trap or after release of the atoms. In order to detect the desired atom configuration, the laser can be flashed to provide fluorescence detection. Additionally, a Stern-Gerlach type magnetic pulse can be applied to resolve the Zeeman levels of the selected hyperfine state. Furthermore, a short pulse of the dipole trap laser can induce delta-kick collimation.\\
While the situation is as described in gravity-assisted measurements, i.e on ground, the situation during operation in microgravity requires a different method since there is no asymmetric potential in lieu of missing gravitational force. One possible approach would be to transfer the atoms to a magnetic-field sensitive spin state and to employ a magnetic field gradient to remove atoms from the optical trap. Ideally, the sequence to create a BEC takes less than 2\,s such that more than 2\,s of free fall time become available for measurements in the Einstein-Elevator, otherwise a scheme with ground loading has to be implemented. 

\section{Outlook}

The INTENTAS apparatus offers flexible capabilities to enable a variety of experimental sequences and operations. The characteristic features, distinguishing it from other compact experiments, are the all-optical approach to BEC creation as well as the emphasis on providing all necessary tools to create an entangled atom source. The combination of the foreseen operation in microgravity and entanglement-enhanced sensing opens up various applications.  

\subsection{Entanglement generation}
\begin{figure}[t]
	\centering
    \includegraphics[width=0.95\columnwidth]{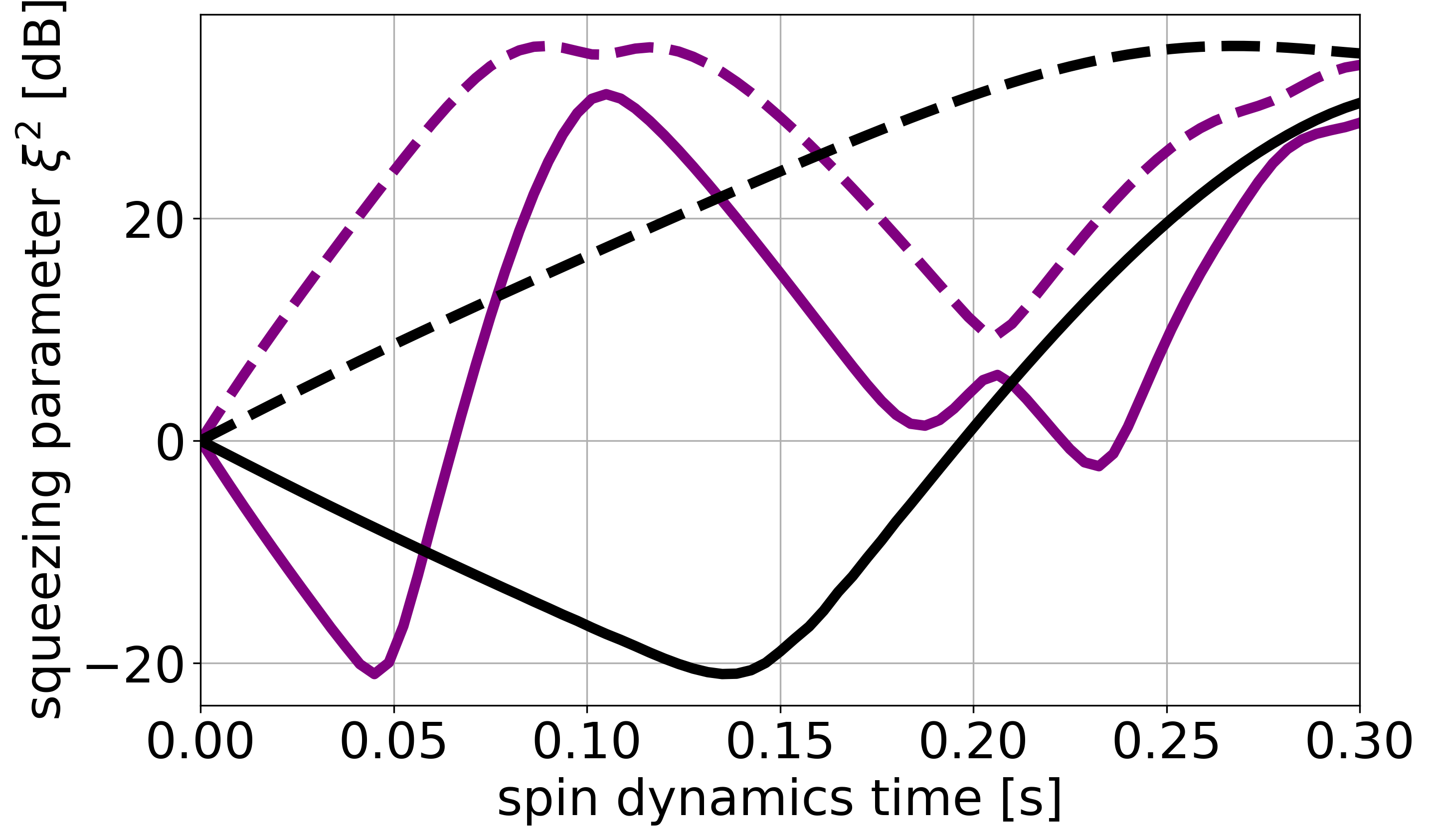}
    \caption{Expected theoretical squeezing parameter $\xi^2$ for 10$^4$ atoms as a function of spin dynamics time for 1\,kHz trap frequency (purple) and 400\,Hz trap frequency (black). The corresponding anti-squeezing in the orthogonal direction is indicated with the dashed lines. The highest squeezing parameter is expected at -21\,dB at 50\,ms and 130\,ms respectively.}
	\label{fig:sq_aq}
\end{figure}
The generation of entanglement can be achieved via different schemes and paths. One procedure to create spin-squeezed states is described in\cite{Cassens2024} and will be used as an example to demonstrate the capabilities of the setup. Here, the entanglement is created using atoms that are initialized in $|F=1,m=0\rangle$. By using microwave dressing, spin-changing collisions are induced transferring atoms into $|F=1,m=\pm 1\rangle$.
Spin-changing collisions generate a two-mode-squeezed vacuum state in the levels $|F=1,m=\pm 1\rangle$. 
Afterwards, the atoms from $|F=1,m=0\rangle$ are transferred to $|F=2,m=0\rangle$.
Subsequently, a $\sigma^-$-polarized rf pulse transfers a single-mode squeezed substate from $|F=1,m=\pm 1\rangle$ to $|F=1,m=0\rangle$.
Together, the single-mode squeezed vacuum state in level $|F=1,m=0\rangle$ and the large number of atoms in level $|F=2,m=0\rangle$ form a two-mode spin-squeezed state.
The remaining atoms in level $|F=1,m=\pm 1\rangle$ are not coupled to the interferometer sequence and can be treated as a negligible loss channel.
A tomography of the spin-squeezed state can be performed by a mw pulse ($\pi$/2) and a subsequent measurement of the population imbalance.
A variation of the mw phase enables a determination of the squeezing angle and the optimal squeezing parameter $\xi^2$ \cite{Pezze18}.
In order to quantify the theoretical limitations of our setup, one can estimate the parameter by solving the respective time-dependent Schrödinger equation using the Crank-Nicolson method \cite{Pezze18}. The result for 10$^4$ atoms is shown in Fig.~\ref{fig:sq_aq}. One can see that, within a reasonable range of trap frequencies, the time that spin dynamics requires to reach the best possible squeezing parameter does not limit the available time for the experimental sequence during the four seconds of free fall. 
Thus, one is able to take advantage of the microgravity environment in order to access extended Ramsey times in combination with entanglement-enhanced sensing. 
We estimate the possible enhancement in a prototypical interferometric measurement, assuming negligible technical noise. Within the exemplary preparation described above, we choose a Ramsey measurement on the Rb clock transition as an example. In order to perform this, two $\pi/2$ pulses with the Ramsey time as delay between them are used after the entangled states are prepared.

\subsection{Exemplary application: Ramsey spectroscopy}
Ramsey spectroscopy on the Rb clock transition not only serves as a secondary definition of the SI second, it also forms the basis for Raman-based atom interferometers that enable the measurement of inertial moments. Therefore, we exemplify the targeted entanglement enhancement by a Ramsey spectroscopy sequence.
In the experimental sequence of Fig.~\ref{fig:Schematic_ES}, we assume the generation of a BEC in level $|F=1, m=0 \rangle$.
The preparation is optionally followed by the generation of squeezing during the "spin dynamics" phase and its transfer to the clock states during the "state manipulation" phase.
During the "interrogation time", we consider the application of two $\pi/2$ pulses on the Rb clock transition with a variable Ramsey time in between.
A final measurement of the number of atoms in the two hyperfine states enables a referencing of the microwave frequency to the atomic hyperfine transition, which forms the basis for a microwave frequency standard.

\begin{figure}[htbp]
	\centering
    \includegraphics[width=0.95\columnwidth]{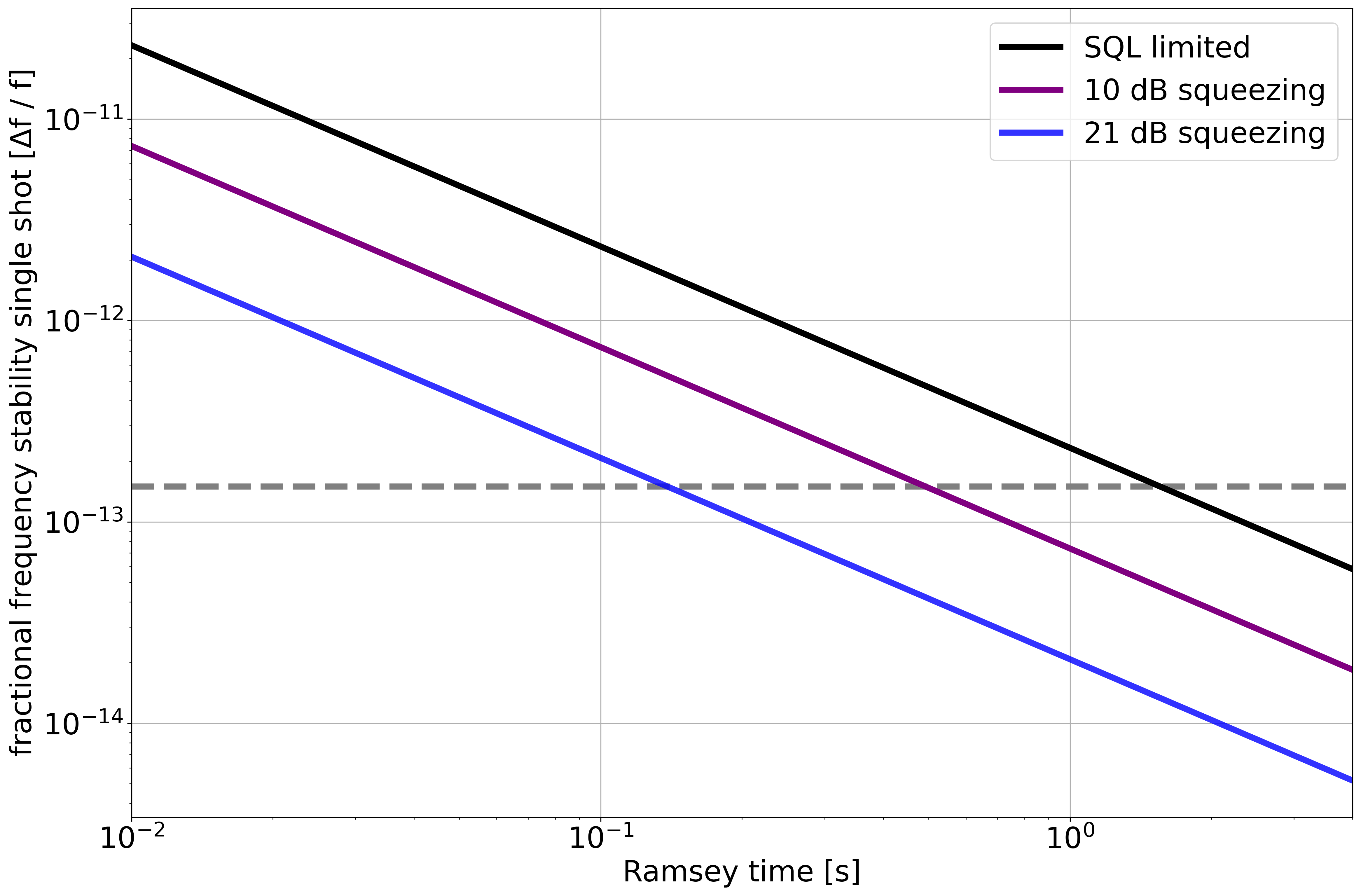}
    \caption{Single shot fractional frequency sensitivity as a function of Ramsey time for a system purely limited by quantum projection noise with 10$^4$ atoms (black), with 10\,dB squeezing (purple) and with the best possible squeezing (blue). Indicated with the gray dashed line is the experimentally achieved sensitivity from a single shot with a comparable setup using cold atoms and a Ramsey time of 5\,s \cite{Szmuk15}.}
	\label{fig:ss_sens}
\end{figure}

Fig.~\ref{fig:ss_sens} shows the single-shot fractional frequency noise for three scenarios, without squeezing, with 10\,dB squeezing that were demonstrated before \cite{Hamley2012}, and with -21\,dB as the theoretical optimum for our apparatus. Close to the optimal case (2\,s Ramsey time and maximum squeezing), a single-shot sensitivity of 1\,$\times$\,10$^{-14}$ would be accessible. However, even with more limited parameters, the experiment can become competitive with state-of-the art compact atomic microwave clocks \cite{Szmuk15},\cite{Laurent2020}. This application demonstrates the advantages of harnessing the unique approach of an entangled compact source with extended Ramsey times in free fall. However, as Ramsey sequences on the Rb clock transition form a basis of both the operation of a microwave frequency standard and inertially sensitive interferometers based on Raman beam splitters, this is only an example and the extension towards the measurement of inertial moments by a Raman-based transfer to finite momentum states~\cite{Anders2021,Cassens2024} is foreseen.

\section*{List of Abbreviations}

\begin{tabular}{ll}
\textbf{Abbreviation} & \textbf{Full Term} \\
AOD  & Acousto-Optical Deflector \\
AOM  & Acousto-Optical Modulator \\
AR   & Anti-Reflective \\
BEC  & Bose-Einstein Condensate \\
DAC  & Digital-to-Analog Converter \\
DDS  & Direct Digital Synthesis \\
ECDL & External Cavity Diode Laser \\
EE   & Einstein-Elevator \\
FPGA & Field-Programmable Gate Array \\
MOT  & Magneto-Optical Trap \\
MW   & Microwave \\
PER  & Polarization Extinction Ratio \\
PID  & Proportional-Integral-Derivative \\
RF   & Radio Frequency \\
SWaP & Size, Weight, and Power \\

\end{tabular}
\section{Declarations}

\subsection*{Ethics approval and consent to participate}
Not applicable

\subsection*{Consent for publication}
Not applicable

\subsection*{Availability of data and materials}
Data are available upon reasonable request. 

\subsection*{Competing interests}
All authors declare that they have no competing interests.

\subsection*{Authors contributions}
A.F. and J.S.H. drafted and edited the manuscript. O.A., A.F., M.G, J.H., J.S.H., A.H., C.K,  M.S. contributed to the writing.  C.K., J.S.H., J.H., A.F. J.K., E.M.R. contributed to the overall concept and design of INTENTAS. O.A., I.B., M.F., M.G., A.H., S.K, M.K, S.K, C.L, K.M., J.P., M.S. S.S., A.W. contributed to designs of subsystems. A.F., C.K., E.G., E.R., W.S., L.W. developed applications scenarios to deduce design requirements. All authors reviewed the manuscript.

\subsection*{Funding}
This work is supported by the German Space Agency (DLR) with funds provided by the BMWK under Grant No. 50WM2174, Grant No. 50WM2175, Grant No. 50WM2176, Grant No. 50WM2177, Grant No. 50WM2178. The authors would also like to thank the DFG and the Lower Saxony state government for their financial support for building the Hannover Institute of Technology (HITec) and the Einstein-Elevator (NI1450004, INST 187/624-1 FUGB) as well as the Institute for Satellite Geodesy and Inertial Sensing of the German Aerospace Center (DLR-SI) for the development and the provision of the carrier system. Supporting work was also contributed by the Deutsche Forschungsgemeinschaft (DFG) under Germany’s Excellence Strategy within the Cluster of Excellence QuantumFrontiers (EXC 2123, Project ID 390837967).

\subsection*{Acknowledgements}
We would like to thank the QUANTUS and MAIUS teams for their insights and support. We are also thankful for the support of Waldemar Herr and Peter Fierlinger with regard to the magnetic shielding. We are also thankful for the support of Ioannis Papadakis with regard to the laser system. The INTENTAS-collaboration also would like to thank the Laboratory of Nano and Quantum Engineering (LNQE) for providing technology and the clean room environment for the assembly of the INTENTAS vacuum system.

\bibliography{intentas_spn.bib} 

\end{document}